\pdfoutput=1
\documentclass[floats,floatfix,amssymb,prd,twocolumn,superscriptaddress,nofootinbib,preprintnumbers]{revtex4-1}

\usepackage{array}[=2016-10-06] 
\usepackage{subcaption,ragged2e}
\DeclareCaptionJustification{justified}{\justifying}
\usepackage{amssymb,amsmath,verbatim,mathtools,needspace,enumitem,etoolbox,graphicx,microtype,afterpage,bm}

\usepackage{tikz}
\usetikzlibrary{shapes.misc, positioning, arrows.meta}
\usepackage{framed}

\usepackage{hyperref}
\hypersetup{colorlinks, citecolor=bluscuro, linkcolor=black, urlcolor=bluscuro}
\definecolor{rossos}{cmyk}{0,1,1,0.55}
\definecolor{bluscuro}{rgb}{0.15, 0.2, .85}
\definecolor{bluchiaro}{cmyk}{1,.3,0.,0.1}
\definecolor{ForestGreen}{rgb}{0.13, 0.55, 0.13}
\definecolor{TLGreen}{RGB}{50, 164, 49}
\definecolor{TLOrange}{RGB}{231,180,22}
\definecolor{TLRed}{RGB}{204,50,50}

\usepackage{multirow,pifont,lmodern,float,tabularx,booktabs}
\usepackage[all]{hypcap}
\usepackage[T1]{fontenc}
\usepackage[utf8]{inputenc}

\renewcommand{\arraystretch}{1.4}

\allowdisplaybreaks

\newcommand{\calP}{{\cal P}}
\newcommand{\be}{\begin{equation}}
\newcommand{\ee}{\end{equation}}
\renewcommand{\d}{{\rm d}}

\newcommand{\unipd}{Dipartimento di Fisica e Astronomia ``G. Galilei'', Universit\`a degli Studi di Padova, via Marzolo 8, I-35131 Padova, Italy}
\newcommand{\infnpd}{INFN, Sezione di Padova, via Marzolo 8, I-35131 Padova, Italy}

\newcommand{\IAP}{Institut d'astrophysique de Paris, UMR 7095 du CNRS et de Sorbonne Universit\'e,\\ 98 bis bd Arago, 75014 Paris, France}

\newcommand{\grappa}{Gravitation Astroparticle Physics Amsterdam (GRAPPA),\\ University of Amsterdam, Science Park 904, 1098 XH Amsterdam, The Netherlands}

\begin{document}

\title{Lattice simulations of scalar-induced gravitational waves from inflation}

\author{Angelo Caravano}
\email{a.caravano@uva.nl}
\affiliation{\grappa}

\author{Gabriele Franciolini}
\email{gabriele.franciolini@unipd.it}
\affiliation{\unipd}
\affiliation{\infnpd}

\author{S\'ebastien Renaux-Petel}
\email{renaux@iap.fr}
\affiliation{\IAP}

\date{\today}

\begin{abstract}
\noindent
  Scalar-induced gravitational waves (SIGWs) provide a powerful probe of inflationary dynamics on scales far smaller than those accessible to the cosmic microwave background and large-scale structure. In scenarios with a transient ultra-slow-roll (USR) phase, the curvature power spectrum can be strongly enhanced on small scales, potentially generating an observable stochastic GW background. In this regime, scalar dynamics during inflation can become nonlinear, challenging the validity of standard perturbative predictions. Existing semi-analytical calculations of SIGWs rely on the linear evolution of inflation fluctuations. In this work, we compute SIGWs from USR inflation using lattice simulations. We evolve the inflaton field nonlinearly during inflation and extract the curvature perturbation nonperturbatively, then simulate its post-reheating horizon re-entry by evolving the Newtonian potential linearly while retaining the full non-Gaussian structure of the initial conditions for the primordial fluctuations in the tensor source. For moderate non-Gaussianity, the semi-analytical prediction captures the correct order of magnitude of the GW signal but receives important corrections. When inflationary non-Gaussianities are large, it can fail dramatically in both amplitude and spectral shape, independently of the overall size of the tensor power spectrum. Our results show that reliable predictions of SIGWs in such scenarios require nonperturbative control of the inflationary scalar dynamics. The code used for this work is available \href{https://github.com/caravangelo/inflation-easy.git}{here}.

\end{abstract}


\maketitle

{
\setcounter{tocdepth}{1}
  \hypersetup{linkcolor=black}
  \tableofcontents
}
\hypersetup{linkcolor=bluscuro}

\section{Introduction}\label{intro}

Stochastic gravitational-wave backgrounds are emerging as a new probe of the early Universe across an enormous range of physical length scales. In particular, gravitational waves (GWs) in the nano--milli-Hz frequency window, targeted by pulsar-timing arrays (PTAs) and future space interferometers, correspond to comoving wavenumbers that are many orders of magnitude smaller than those constrained by the cosmic microwave background (CMB) and large-scale structure (LSS). They therefore probe inflationary fluctuations on scales that are essentially unconstrained by standard cosmological observations. Recent PTA data sets already provide evidence for a nanohertz stochastic background~\cite{NANOGrav:2023gor,EPTA:2023fyk}, whose nature remains unknown and may be astrophysical or cosmological (see e.g. \cite{Sesana:2025udx} for a review). Planned observatories such as LISA~\cite{LISA:2024hlh}, DECIGO~\cite{Kawamura:2020pcg}, Cosmic Explorer~\cite{Evans:2021gyd} and the Einstein Telescope ~\cite{ET:2019dnz,ET:2025xjr} will dramatically extend the accessible frequency range and sensitivity. 

A particularly well-motivated target is the stochastic background of scalar-induced gravitational waves (SIGWs), tensor modes generated at second order by amplified scalar perturbations~\cite{Tomita:1975kj, Matarrese:1992rp,Matarrese:1993zf,Matarrese:1997ay, Acquaviva:2002ud, Mollerach:2003nq,Carbone:2004iv, Ananda:2006af, Baumann:2007zm} (for a review, see~\cite{Domenech:2021ztg}). In many inflationary scenarios that enhance the curvature power spectrum on small scales, often motivated by primordial black hole (PBH) phenomenology~\cite{Byrnes:2025tji}, the associated SIGW signal can fall in the PTA or LISA bands and can become observationally relevant (see e.g.~\cite{Fumagalli:2021dtd,LISACosmologyWorkingGroup:2024hsc,LISACosmologyWorkingGroup:2025vdz} for associated LISA forecasts). SIGWs therefore provide a direct avenue to probe inflationary physics on scales far beyond those accessible to the CMB and LSS.

Despite the maturity of the SIGW framework, existing predictions rely on a set of approximations that can become uncontrolled precisely in the regimes of greatest interest. The standard approach computes SIGWs at second order in perturbation theory while treating the scalar sector in the Gaussian approximation and assuming linear evolution of perturbations during inflation and at Hubble re-entry. This leads to semi-analytical expressions for $\Omega_{\rm GW}(k)$ in terms of the scalar power spectrum (see Refs.~\cite{Espinosa:2018eve,Kohri:2018awv} and references therein).
This strategy is powerful and efficient, but it is not designed to capture genuinely nonperturbative scalar dynamics during inflation, nor the impact of large primordial non-Gaussianities on the post-reheating emission. A growing literature has explored non-Gaussian corrections perturbatively in specific templates, for example local-type non-Gaussianity, and in diagrammatic frameworks, see e.g.~\cite{Cai:2018dig,Unal:2018yaa,Yuan:2020iwf,Atal:2021jyo,Adshead:2021hnm,Garcia-Saenz:2022tzu,Yuan:2023ofl,Li:2023qua,Li:2023xtl,Perna:2024ehx,Ruiz:2024weh,Li:2025met}. These studies notably highlight that non-Gaussianity can modify both the amplitude and the shape of the SIGW spectrum. However, once the scalar sector becomes strongly nonlinear, it is not obvious that any truncation of the non-Gaussian expansion remains reliable.\footnote{The perturbative regime is typically violated when considering PBH formation (see e.g. \cite{Ferrante:2022mui}). Even though SIGWs are not specifically sensitive to the tail of the density perturbation distribution, and thus one expects milder effects, NGs can still play a crucial role in determining the shape of the observed spectrum. }

In this work, we go beyond these standard approximations using lattice simulations. It is a follow-up of our previous studies of ultra-slow-roll (USR) inflation on the lattice~\cite{Caravano:2024moy,Caravano:2025diq}. In these papers we developed and validated a framework to simulate the {nonlinear} dynamics of the inflaton during SR to USR to SR transitions. We extracted the curvature perturbation $\zeta$ using a nonperturbative $\delta N$ procedure, thereby capturing both intrinsic non-Gaussianity generated around horizon exit by nonlinear field dynamics, and the nonlinear mapping between the inflaton configuration and $\zeta$~\cite{Caravano:2025diq}. The present paper extends that program to the tensor sector. We compute the SIGW background sourced by these non-Gaussian primordial perturbations and provide a first-principles benchmark for when and how the standard semi-analytical treatment breaks down.

\begin{figure*}
    \centering
    \includegraphics[width=0.9\linewidth]{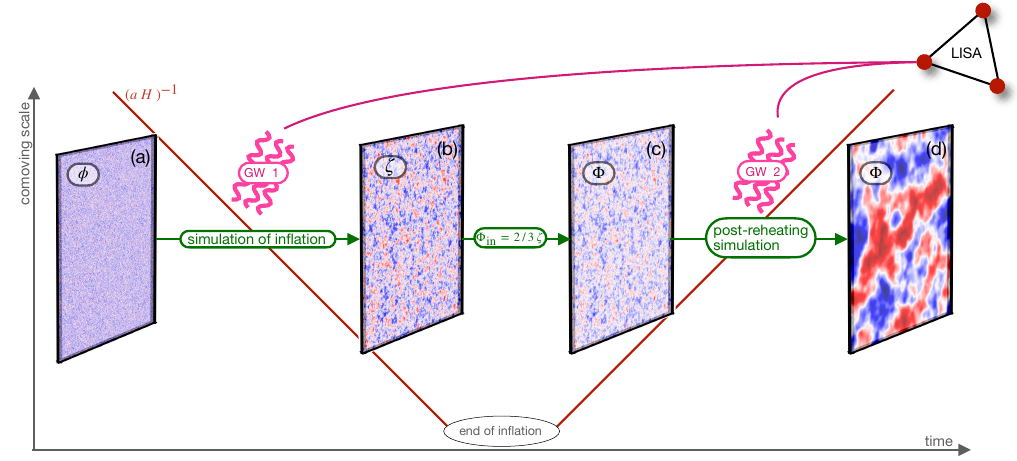}
    \caption{\textit{Simulation pipeline:} From sub-horizon inflaton fluctuations (a), we obtain the super-horizon curvature perturbation (b) using the lattice simulation of inflation. This curvature field is then mapped to the super-horizon Newtonian potential after inflation (c), which sets the initial conditions for the simulation of horizon re-entry. Evolving this system yields the sub-horizon Newtonian potential (d). The total gravitational-wave signal contains two contributions: the inflationary component (GW~1), emitted at horizon exit, and the post-reheating component generated at horizon re-entry (GW~2).~\footnote{In general, there is also a mixed component coming from the correlation between the two. The post-reheating component is anyway largely dominant in the scenarios considered here.} 
The GWs would then travel freely to reach late time observers, like LISA,  shown as an example.
    Figure adapted from \cite{Caravano:2025klk}.}
    \label{fig:pipeline}
\end{figure*}

Concretely, our calculation proceeds in two stages, summarized in Fig.~\ref{fig:pipeline}. First, we perform a nonlinear lattice simulation of inflationary scalar-field dynamics, starting from sub-horizon fluctuations and evolving through the USR phase. From the lattice output we construct the frozen super-horizon curvature perturbation $\zeta(\bm x)$ nonperturbatively. Second, we propagate these primordial perturbations into the post-reheating era and simulate horizon re-entry on a lattice. We evolve the Newtonian potential $\Phi$ with its linear equation of motion in a radiation-dominated background while retaining the full nonlinear structure of the initial conditions. In this post-reheating stage we compute SIGWs sourced by the standard second-order scalar stress, without imposing the Gaussian approximation. All primordial higher-point information is therefore included implicitly through the real-space fields inherited from inflation. In addition, we evolve the tensor modes during inflation to quantify the inflationary contribution to the total signal in our scenarios. The code used for this work is publicly available at the following \href{https://github.com/caravangelo/inflation-easy.git}{link}

The paper is organized as follows. In Sec.~\ref{sec:analytical} we review the standard semi-analytical SIGW calculation and fix conventions. In Sec.~\ref{sec:lattice} we describe our lattice framework for evolving both the scalar sector and the induced tensor modes, including our treatment of transverse-traceless projection and GW observables. Sec.~\ref{sec:models} presents the inflationary potentials considered in this work. Our numerical results are given in Sec.~\ref{sec:results}, where we compare directly to semi-analytical predictions. In Sec.~\ref{sec:trapping}, we discuss the trapping phenomenon and the associated multi-peak structure that emerge in the large non-Gaussianity regime. We conclude in Sec.~\ref{sec:conclusions}.

\paragraph*{Notation}
Throughout this paper, we describe the metric as a perturbation of the FLRW metric in the longitudinal, or conformal Newtonian, gauge
\begin{equation}
\label{eq:metric newtonian}
\begin{aligned}
\d s^2 ={}& -a^2(1+2\Phi^{(1)}) \d \eta^2 \\
&+ a^2 \left[ (1-2\Phi^{(1)})\delta_{ij}
+\frac{1}{2}h_{ij}^{(2)} \right] \d x^i \d x^j \, .
\end{aligned}
\end{equation}
where $\eta$ is the conformal time. We neglect vector perturbations and anisotropic stress, and therefore identify
the two scalar Bardeen potentials, $\Phi=\Psi$. The regime of
validity of these approximations is discussed in
Sec.~\ref{subsec:lattice-postinflation}. The factor $1/2$ follows the standard convention in the SIGW literature when defining second-order tensor modes. We assume that linear tensor perturbations are negligible. We will drop the superscripts denoting perturbative order when no confusion can arise.

\section{Analytical estimation of SIGWs}
\label{sec:analytical}
The derivation of second-order induced GWs follows\,\cite{Tomita:1975kj,Matarrese:1993zf,Acquaviva:2002ud,Mollerach:2003nq,Ananda:2006af,Baumann:2007zm}.
In the metric defined in Eq.~\eqref{eq:metric newtonian}, we ignore tensor perturbations generated at first order $h_{ij}^{(1)}$ and consider scalar perturbations that act as a source of second order tensor modes at  $h_{ij}^{(2)}$. The exact evolution of $h_{ij}^{(2)}$ can be obtained from the spatial part of the Einstein equations after applying the projection tensor $\mathcal T_{ij}^{lm}$, which selects the transverse-traceless component.
In the absence of anisotropic stress, at second order (note that we drop the superscript indicating the order in perturbation theory from now on) one obtains \cite{Acquaviva:2002ud, Mollerach:2003nq, Ananda:2006af, Baumann:2007zm}
\be
h_{ij}''(\eta,{\bm x})
+2\mathcal H h_{ij}'(\eta,{\bm x})
-\nabla^2 h_{ij}(\eta,{\bm x})
=-4 \mathcal T_{ij}{}^{lm}\mathcal S_{lm}(\eta,{\bm x}),
\label{eq:eom GW}
\ee
where $'$ is the derivative with respect to conformal time $\eta$, $\mathcal H=a'/a$ denotes the conformal Hubble parameter,
$S_{ij}$ is the source term.
The transverse and traceless projector reads 
\begin{align}
\mathcal T_{ij}{}^{lm} &=
 \mathcal T_{(i}{}^l \mathcal T_{j)}{}^m
- \frac12 \mathcal T_{ij} \mathcal T^{lm},
\nonumber \\
\mathcal T_{ij} &= \delta_{ij} - \frac{\partial_i \partial_j}{\partial^2},
\label{eq:PTT}
\end{align}
where spatial indices are raised and lowered with $\delta_{ij}$.

Solving Eq.~\eqref{eq:eom GW} in Fourier space, we obtain
\begin{align}
\big\langle
h_{ij}(\eta,\bm k)\,
h^{ij}(\eta,\bm k')
\big\rangle
&= (2\pi)^3\delta^{(3)}(\bm k+\bm k') \, 2P_h(\eta,k)\,,
\label{eq:-Power-spectrum-def}
\end{align}
from which one deduces the dimensionless power spectrum $ \calP_h (\eta_f,k)=\frac{k^3}{2\pi^2}P_h(\eta_f,k)$ of GWs at a time $\eta_f$ after the end of their production. The fractional energy density of GWs per logarithmic interval in frequency is then given by \cite{Maggiore:1999vm}
\begin{equation}
\Omega_{\rm GW}(\eta_f, k)\equiv
  \frac{\rho_{\rm GW}(\eta_f, k)}{\rho_{\rm c}(\eta_f)}
  = \frac{1}{24} \left (\frac{k}{\cal H(\eta)}\right )^2 \overline{\calP_h(\eta_f, k)} \,,
\label{eq:OmegaGW}
\end{equation}
where the bar denotes a time average over multiple oscillations and $\rho_{\rm c}$ is the critical energy density.

Most of the emission is expected to take place when GW modes cross the Hubble sphere, a condition which can be met during inflation and after reheating in the subsequent cosmological evolution. In the models considered in this work, the dominant contribution comes from the post-reheating emission at horizon re-entry. For this reason, the analytical estimates presented below focus on the post-reheating signal. The inflationary contribution is included in our numerical pipeline (see Appendix~\ref{app:inflationary-gws}), but a consistent analytical treatment requires additional work to renormalize the calculation, and is left for future work.

\subsection{Post-reheating contribution}

We assume that, after reheating, the background is described by an FLRW universe filled with a perfect fluid with constant equation-of-state parameter $w$ and sound speed $c_s^2=w$, and we neglect anisotropic stress at first order so that $\Phi=\Psi$. Scalar perturbations are then governed by the (linear) Bardeen equation (Eq.~(8.31) in Ref.~\cite{Malik:2008im})
\begin{align}\label{eq:phievo_full}
{\Phi}''(\eta,{\bm x})
&+3{\cal H}(1+c_s^2){\Phi}'(\eta,{\bm x})
\nonumber \\
&+
\left[ 2 {\cal H}'
+(1+3 c_s^2){\cal H}^2
- c_s^2\nabla^2 \right]
{\Phi}(\eta,{\bm x})=0,
\end{align}
which, for $c_s^2=w=\mathrm{const}$, reduces to
\begin{align}\label{eq:phievo}
{\Phi}''(\eta,{\bm x})
&+3{\cal H}(1+c_s^2){\Phi}'(\eta,{\bm x})
- c_s^2\nabla^2
{\Phi}(\eta,{\bm x})=0.
\end{align}
Throughout the remainder of this work, we will assume perfect radiation domination (RD)
and set $c_s^2= w = 1/3$. However, we mention that generalizing the result to different thermal histories (such as the QCD transition) fits within the same framework by allowing time-varying equation-of-state parameters.

We decompose $\Phi_{\vec k}(\eta)$ into the transfer function $T_\Phi(k,\eta)$ and the primordial perturbation $\Phi_{0,\vec k}$ as
\begin{equation}
\Phi_{\vec k}(\eta) = T_\Phi(k,\eta)\,\Phi_{0,\vec k}.
\end{equation}
The primordial perturbation is related to the super-Hubble curvature perturbation on uniform energy density hypersurfaces $\zeta_{\vec k}$ via
\begin{equation}
\Phi_{0,\vec k} = \frac{2}{3}\,\zeta_{\vec k}
\label{pert}
\end{equation}
and the transfer function reads
\begin{align}
 T_\Phi(k,\eta) &= 3
 \left [
\frac{\sin(k\eta/\sqrt{3}) - (k\eta/\sqrt{3})\cos(k\eta/\sqrt{3})}{(k\eta/\sqrt{3})^3}
\right]\,
 \end{align}
 where $k \eta \to 0$ deep in the super-Hubble regime. From $\Phi$, one can then deduce the source of GW in the Newtonian gauge, see Eq.~\eqref{eq:eom GW}, which takes the form
\begin{align}
{\cal S}_{ij} (\eta,{\bm x})
&= 4\Phi\partial_i\partial_j\Phi+2\partial_i\Phi\partial_j\Phi
\nonumber \\
&-\frac{4}{3(1+w)}\partial_i\left(\frac{\Phi'}{\mathcal H}+\Phi\right)\partial_j\left(\frac{\Phi'}{\mathcal H}+\Phi\right)\,.
\label{eq:source S}
\end{align}

\subsection{Perturbation theory prediction}
\label{semi-analytical-GW-spectrum}
Assuming curvature perturbations are Gaussian, the leading-order prediction for the tensor power spectrum emitted after reheating is
\begin{multline}
\overline{\mathcal{P}_h (\eta, k)} = 4
\int_0^\infty \text{d}t \int_{0}^{1}\text{d} s \left [ \frac{t(2+t)(1-s^2)}{
(1-s+t)(1+s+t)
} \right ]^2
\\
\times
\overline{I^2 (t,s,k,\eta)} \mathcal{P}_\zeta \left(k\,
u\right) \mathcal{P}_\zeta \left(k\,
v \right) ,
\label{eq:P_h_ts}
\end{multline}
where $ u \equiv \frac{t+s+1}{2} $ and $v \equiv \frac{t-s+1}{2} $.
When the emission takes place in a RD universe, the kernel function in the deep sub-horizon regime $k\eta \to \infty$ takes the form\,\cite{Espinosa:2018eve,Kohri:2018awv}
\begin{align}
&\overline{I_{\text{RD}}^2(t, s, k\eta \to \infty)} =\frac{1}{2 (k \eta)^2}
I_A^2(u,v)\left[I_B^2(u,v)+I_C^2(u,v)\right]
\label{I_RD_osc_ave_ts}
\end{align}
where
\begin{equation}
\begin{aligned}
I_A(u,v)&= \frac{3(u^2+v^2-3)}{4u^3v^3} \,,\\
I_B(u,v)&= -4uv+(u^2+v^2-3)\log\left|\frac{3-(u+v)^2}{3-(u-v)^2}\right| \,,\\
I_C(u,v)&= \pi(u^2+v^2-3)\Theta(u+v-\sqrt{3}) \,,
\label{IABC-functions}
\end{aligned}
\end{equation}
$u$ and $v$ have been introduced above and
$\Theta(x)$ is the Heaviside function.

\section{Lattice calculation of SIGWs}
\label{sec:lattice}

In this section we describe the lattice computation of SIGWs used in this work. The simulation code is publicly available and can be found in~\cite{Caravano:2025klk}.
The simulation proceeds in two stages:
(i) a {nonlinear} evolution of the inflaton during inflation, and
(ii) a post-reheating evolution of scalar perturbations in terms of the Newtonian potential in a radiation-dominated universe.
In both stages we evolve the six independent components of the tensor perturbation $h_{ij}$ and apply the transverse--traceless (TT) projection only when constructing gravitational-wave observables. In the main text, we focus on the post-reheating contribution generated at horizon re-entry, which dominates in the scenarios considered in this work. The inflationary contribution and its numerical implementation are described in Appendix~\ref{app:inflationary-gws}.
\subsection{Gravitational wave evolution, TT projection, and power spectrum}
\label{subsec:lattice-gw-eq}

In the analytical calculation, the second-order tensor perturbations satisfy Eq.~\eqref{eq:eom GW}, where the TT projector is given in Eq.~\eqref{eq:PTT} and the scalar source term is ${\cal S}_{ij}$ (for the post-reheating contribution, Eq.~\eqref{eq:source S}).

On the lattice, we evolve the six independent components of a symmetric tensor
$h_{ij}$ without imposing transversality or tracelessness during the evolution. They obey
\begin{equation}
h_{ij}'' + 2\mathcal H h_{ij}' - \nabla^2 h_{ij}
= -4\,{\cal S}_{ij},
\label{eq:lattice-eom-hij}
\end{equation}
where ${\cal S}_{ij}$ is evaluated either from the inflaton (during the inflationary
stage) or from the Newtonian potential (during the post-reheating stage).\footnote{
While we work in a specific gauge, it has been shown that the abundance of late-time SIGWs emitted during a RD phase is gauge-independent~\cite{DeLuca:2019ufz,Inomata:2019yww,Yuan:2019fwv,Domenech:2020xin} and coincides with that computed in the Newtonian gauge when neglecting the back-reaction of GWs on the spacetime metric~\cite{Domenech:2025ccu}.
}
The TT projection is not applied during the evolution, but only when computing gravitational-wave observables, as commonly done in the literature.
Since the TT projector commutes with the linear operator on the left-hand side of Eq.~\eqref{eq:lattice-eom-hij}, this procedure is equivalent to evolving
the TT modes directly~\cite{Garcia-Bellido:2007fiu}.

For each Fourier wavevector we construct the TT component through Eq.~\eqref{eq:PTT}, which, in momentum space, takes the form:
\begin{equation}
P_{ij}=\delta_{ij}-\hat{k}_i\hat{k}_j, \qquad
\Lambda_{ij}{}^{l m}
= P_{(i}{}^l P_{j)}{}^m - \tfrac12\,P_{ij}P^{l m},
\end{equation}
so that
\begin{equation}
h_{ij}^{\rm TT}(\eta,\bm k)
= \Lambda_{ij}{}^{l m}(\hat{\bm k})\,h_{l m}(\eta,\bm k).
\end{equation}
The tensor power spectrum is computed following \eqref{eq:-Power-spectrum-def}, where modes are assigned to spherical shells in Fourier space using the mapping between lattice momenta and physical wavenumbers introduced in \cite{Caravano:2021pgc, Caravano:2022yyv}. One then deduces $\Omega_{\rm GW}$ as given in \eqref{eq:OmegaGW}. Notice that the time-average is not needed on the lattice, replaced by the average over lattice points of the same wavenumber. Eventually, for the post-reheating contribution, and as the GW energy density scales like the one of radiation, we can compute the present-day spectrum using 
\begin{equation}
\Omega_{\rm GW,0}(k)
= \Omega_{r,0}\,\Omega_{\rm GW}(\eta_f,k),
\label{eq:latticeOmega0}
\end{equation}
with $\eta_f$ chosen sufficiently deep in the radiation era so that all relevant emission processes have occurred. In the last equation, $\Omega_{r,0}$ is the current radiation abundance.

\subsection{Post-reheating evolution}
\label{subsec:lattice-postinflation}

After inflation, we evolve the Newtonian potential $\Phi$ using the Bardeen equation given in Eq.~\eqref{eq:phievo} for a perfect fluid with constant equation-of-state parameter $w=c_s^2$. The relevant second-order source of induced GWs is given by Eq.~\eqref{eq:source S}.
This expression is evaluated nonperturbatively in real space using the fields $\Phi(\eta,\bm x)$ and $\Phi'(\eta,\bm x)$ evolved on the lattice, while the super-Hubble initial condition is fixed by the curvature perturbation extracted from the inflationary simulation and mapped to $\Phi$ through Eq.~\eqref{pert}.
This post-reheating evolution follows the same physical equations as the standard calculation reviewed in Sec.~\ref{sec:analytical} (and, in particular, it relies on the linear scalar dynamics), but it retains the full non-Gaussian structure of the primordial initial conditions inherited from the nonlinear lattice simulation of the inflationary scalar sector.~\footnote{The post-inflation evolution is equivalent to the one used in~\cite{Zeng:2025cer}, but with the important difference that our approach uses initial conditions coming from the nonperturbative inflationary simulation.}

The tensor modes obey Eq.~\eqref{eq:lattice-eom-hij} with ${\cal S}_{ij}$ given by Eq.~\eqref{eq:source S}.
At the end of the simulation, the TT projection is applied to the tensor field in Fourier space, and the corresponding fractional energy density is computed using Eq.~\eqref{eq:latticeOmega0}.

A limitation of our approach is that we evolve the gravitational potential linearly; vector/shear perturbations and anisotropic stress are also neglected. However, it is interesting to notice that the radiation-era second-order transfer function scales in time exactly as the linear one~\cite{DeLuca:2023tun}. As a result, nonlinear corrections are suppressed by at least one power of the gravitational potential and are expected to remain small as long as $|\Phi|\ll 1$. Vector/shear perturbations and anisotropic-stress contributions are likewise expected to remain subleading in this regime, but may become relevant when $|\Phi|$ approaches unity, as can occur locally in the most extreme scenarios discussed in Sec.~\ref{sec:trapping}. In that situation, one should also take into account
the backreaction of inhomogeneities on the background evolution, for instance
through relativistic hydrodynamical simulations; see, e.g.,~\cite{Ning:2025yvj} for progress in that direction.

\section{Inflationary models}
\label{sec:models}

In this section, we present the inflationary models considered in this work. We use a representative toy model of single-field inflation featuring a transient phase of USR. The inflaton field $\phi$ is minimally coupled to gravity, and the dynamics of the model is fully specified by the inflaton potential $V(\phi)$.

Following Ref.~\cite{Franciolini:2022pav} (see also Refs.~\cite{Byrnes:2018txb,Taoso:2021uvl}), we model the background evolution as an initial SR phase, during which the inflaton smoothly rolls down its potential and generates CMB-scale perturbations. 
In a subsequent phase, as the field approaches an approximate inflection point, the system enters a transient USR regime characterized by $
{\eta_H} \equiv -\ddot H/({2 H \dot H}) \gtrsim 3/2$, during which curvature perturbations grow rapidly on super-Hubble scales.
In the stated condition, $H$ is the Hubble parameter, overdots indicating derivatives with respect to cosmic time $t$, and ${\eta_H}$ is the second {Hubble} SR parameter.  
The USR phase is subsequently followed by a second SR stage with a negative value of ${\eta_H}$. Throughout the evolution, the first slow-roll parameter $\epsilon \equiv  - \dot H/H^2$ remains small, ensuring an approximately quasi-de Sitter background. This behavior is implemented by prescribing a three-stage piecewise-constant profile for ${\eta_H(N)}$, with smooth interpolations between the phases: an initial SR stage with ${\eta_{H,{\rm I}}} \ll 1$, a transient USR stage with ${\eta_{H,{\rm II}}} > 3/2$, and a final SR phase with ${\eta_{H,{\rm III}}} \leq 0$. Given this choice of ${\eta_H(N)}$, the corresponding inflaton potential can be reconstructed following the procedure detailed in Ref.~\cite{Franciolini:2022pav}.
To explore scenarios with different amplitudes of the curvature power spectrum and distinct relevance of the primordial non-Gaussianity, we vary both the duration of the USR phase in terms of the number of e-folds,
$\Delta N \equiv N_{\rm end} - N_{\rm in}$,
and the values of ${\eta_{H,{\rm II}}}$ during USR and ${\eta_{H,{\rm III}}}$ after USR. Since super-Hubble perturbations grow exponentially with $N$ during USR, longer USR phases result in a stronger enhancement of the power spectrum and in larger non-Gaussian signatures, as we will see. For a detailed description of the constructed potential, along with illustrative plots, see Sec.~II of Ref.~\cite{Caravano:2024moy}.

The lattice simulation of the USR phase allows us to track in a {nonperturbative} manner the dynamics of the inflaton field $\phi$. 
In order to compute post-reheating observables, it is however necessary to reconstruct the curvature perturbation $\zeta$, which freezes on super-Hubble scales after the end of the USR epoch. The relation between $\zeta$ and the inflaton fluctuations is intrinsically non-linear. As discussed in detail in \cite{Caravano:2025diq}, this mapping can be non-perturbatively derived by using the $\delta N$ formalism applied to the inflaton configuration obtained from the simulation. In the regime of small fluctuations, the resulting expression reproduces the commonly adopted logarithmic relation $\zeta = \frac{1}{{\eta_{H,{\rm III}}}}\log\!\left(1+{\eta_{H,{\rm III}}}\, \zeta_{\rm lin}\right)$ (see, e.g., \cite{Cai:2018dkf,Atal:2019cdz,Biagetti:2021eep, Pi:2022ysn,Ballesteros:2024pwn,Inui:2024fgk}), while significant deviations arise in the non-perturbative and/or trapping regime. In what follows, we will therefore employ {the nonperturbatively reconstructed curvature perturbation} $\zeta$ in order to derive our results and consistently assess the impact of the induced non-Gaussianity on the SIGW spectrum.

We divide the models into two main categories. The first is the one in which non-Gaussianity is mild. These are the exact models studied in \cite{Caravano:2024moy, Caravano:2025diq}, with three sub-categories depending on the relevance and sign of the inflationary self-interaction during the USR period. The second category is the one in which non-Gaussianity is large, which will lead to much larger deviations from the analytical results, as we will see.

\subsection{Mild non-Gaussianity}

 The specific parameter choices adopted in our analysis are summarized in Tab.~\ref{tab:usr_cases} (see Ref.~\cite{Caravano:2024moy} for more details). 
 The classification is made depending on the sign of the inflaton leading order (i.e. cubic) self-interaction.
It is worth emphasizing that, for constant ${\eta_H}$, the second derivative of the reconstructed potential is invariant under the transformation ${\eta_H} \rightarrow 3 - {\eta_H}$, a symmetry known as Wands duality~\cite{Wands:1998yp}. 
If the transition between USR and the subsequent SR phase takes place between dual phases, and the transition is smooth, higher-order derivatives of the potential are suppressed, and inflaton self-interactions during USR are consequently negligible. On the other hand, if ${\eta_{H,{\rm II}}} \neq 3-{\eta_{H,{\rm III}}}$, inflaton self-interaction can be large. 

Depending on the value of ${\eta_H}$ during the USR phase, we distinguish three qualitatively different scenarios, summarized in Tab.~\ref{tab:usr_cases}:
\begin{itemize}[noitemsep,left=0pt]
    \item \textbf{Case I: Wands-dual regime.}  
    In this case, the effective mass parameter $\nu^2 \simeq 9/4-V''/H^2$, remains approximately constant after the onset of USR. Inflaton self-interactions are strongly suppressed, and the dynamics of the field fluctuations $\delta\phi$ are well approximated by a nearly free theory. In this case, the whole curvature non-Gaussianity is induced by the GR-induced non-linearity relating $\delta \phi$ and $\zeta$.

    \item \textbf{Case II: Repulsive interactions.}  
    Here, $\nu^2$ evolves significantly during USR, increasing near the end of the phase (around $N - N_{\rm in} \sim 2$). This behavior induces a positive cubic coupling, corresponding to repulsive self-interactions. As a consequence, the resulting non-Gaussianities suppress large positive excursions of the inflaton perturbations.

    \item \textbf{Case III: Attractive interactions.}  
    In contrast, this scenario features a decreasing $\nu^2$ close to the USR-to-SR transition, leading to a negative cubic coupling and attractive self-interactions. The associated non-Gaussianities enhance large positive inflaton fluctuations.
\end{itemize}

In order to investigate the impact of non-Gaussianities and non-linear effects on the generation of SIGWs, we consider scenarios in which the amplitude of the curvature perturbation is enhanced at different levels relative to the values observed at CMB scales. This enhancement is controlled by varying the USR duration, $\Delta N$, while keeping all other parameters, as well as the structural features defining cases I, II, and III, fixed. Increasing $\Delta N$ results in a larger amplification of the curvature perturbations, as summarized in Tab.~\ref{tab:usr_cases}.

The reconstructed potentials are monotonous in Case~II, while in Cases~I and III they exhibit a local maximum and a local minimum. In the latter situations, regions of the Universe may become trapped in the local minimum, potentially leading to primordial black hole formation right after inflation. These aspects, together with the properties of super-Hubble inflaton perturbations at the end of the USR phase, and their implications for late-time observables, are discussed in detail in Ref.~\cite{Caravano:2024moy}. 

In the following, we normalize the momenta with respect to $k_{\rm ref}$, which is the mode crossing the Hubble scale at $N_{\rm in}$, i.e. $k_{\rm ref} = a(N_{\rm in}) H(N_{\rm in})$.

{
\renewcommand{\arraystretch}{1.4}
\setlength{\tabcolsep}{3pt}
\begin{table}
\caption{ Parameters adopted in the various scenarios considered in this work. The value of $\Delta N$ is given as a function of the maximum spectral amplitude ${\cal P}_{\zeta,{\rm tree}}^{\rm max}$ obtained with the tree-level computation. For the large-NG case, we fix the duration of the CR phase to 1 e-fold.}
\begin{tabularx}{1 \columnwidth}{|X|c|c|c|}
\hline
\hline
& ${\eta_{H,{\rm II}}}$ & ${\eta_{H,{\rm III}}}$ & $\Delta N \equiv N_{\rm  end} - N_{\rm  in}$ 
\\
\hline
\hline
\multicolumn{4}{|c|}{Mild-NGs} \\
\hline
\hline
Case I (Wands duality) & $3.5$ & $-0.5$ &
$2.6+ 
0.29 \log_{10} {\cal P}_{\zeta,{\rm tree}}^{\rm max} 
$
\\
\hline
Case II (repulsive) & $3$& $-0.5$ &
$3.3+ 
0.38 
\log_{10} {\cal P}_{\zeta,{\rm tree}}^{\rm max} 
$
\\
\hline
Case III (attractive)  &$4.5$ & $-0.5$ & 
$1.8+ 
0.19 \log_{10} {\cal P}_{\zeta,{\rm tree}}^{\rm max} 
$
\\
\hline
\hline
\multicolumn{4}{|c|}{Large-NGs} \\
\hline \hline 
USR-CR-SR
& $8$ & $-5$ &
$0.86 + 0.088
\log_{10} {\cal P}_{\zeta,{\rm tree}}^{\rm max} 
$
\\
\hline
\hline
\end{tabularx}
\label{tab:usr_cases}
\end{table}
}

\subsection{Large non-Gaussianity}
\label{large-NG}

In this section, we describe the class of scenarios characterized by large non-Gaussianity that we consider. In order to enhance non-Gaussian effects, we consider configurations in which ${\eta_{H,{\rm III}}}$ takes significantly larger (and negative) values compared to the mild-NG cases. In this regime, both sources of non-Gaussianity become more prominent: on the one hand, inflaton self-interactions are enhanced due to the rapid evolution of the background, while on the other hand the non-linear relation between $\delta\phi$ and $\zeta$, encoded in the approximate logarithmic mapping, induces stronger GR non-linearities.

In contrast to the mild-NG case, we do not further subdivide these scenarios into distinct subcases. This is because achieving a Wands-dual configuration requires very large values of ${\eta_{H,{\rm II}}}$ in order to satisfy the condition ${\eta_{H,{\rm II}}} = 3 - {\eta_{H,{\rm III}}}$. Such large values inevitably lead to very sharp transitions into and out of the USR phase. Recalling that the effective mass parameter takes the form $\nu^2 \simeq 9/4 - \left[ {\eta_H} (3-{\eta_H}) + \d {\eta_H}/\d N \right ]$, it is easy to see that rapid ${\eta_H}$ variation inevitably induces variations of the effective mass parameter. 
These transitions, in turn, generate large higher-order derivatives of the potential, resulting in strong inflaton self-interactions independently of the precise choice of parameters. As a consequence, all realizations in this class share qualitatively similar features, characterized by large intrinsic and induced non-Gaussianities.

Furthermore, in order to sustain inflation for a sufficient number of e-folds and allow the relevant modes to freeze out, we extend the background template by introducing a fourth stage following the end of phase III. In this additional phase, we set ${\eta_{H,{\rm IV}}} \simeq 0$, which prevents the first slow-roll parameter $\epsilon$ from growing to $\mathcal{O}(1)$ too rapidly and prematurely ending inflation. This modification ensures a controlled evolution after the USR phase while preserving the strong non-linear effects accumulated during it for the relevant modes. We note that the intermediate stage characterized by a large and negative value of ${\eta_{H,{\rm III}}}$, during which $\epsilon$ remains small but grows steadily, corresponds to a constant-roll (CR) phase. In this regime, the second slow-roll parameter is constant and large in magnitude, leading to a departure from standard SR dynamics. Upon entering stage IV, where ${\eta_{H,{\rm IV}}} \simeq 0$, the system relaxes back to the usual SR attractor, ensuring that inflation can proceed in a standard way until its eventual termination.

As in the mild-NG scenarios, we explore different amplitudes of the curvature power spectrum by varying the duration of the USR phase, $\Delta N$, while keeping the remaining parameters fixed. Longer USR phases correspond to larger enhancements of ${\cal P}_\zeta$, and therefore to increasingly strong non-Gaussian signatures. The parameters chosen in this case (denoted USR-CR-SR) are shown in Tab.~\ref{tab:usr_cases}.

\section{Simulation results}
\label{sec:results}

In this section we present the GW spectra obtained with the lattice method introduced above.

\subsection{Mild non-Gaussianity}
\label{sec:results_mild}

For the cases with mild NG, the scalar dynamics have already been investigated in detail in \cite{Caravano:2024moy,Caravano:2025diq}. In all these models, the inflationary contribution to the GW spectrum (the component emitted during inflation) is so small that it lies below the numerical noise and cannot be reliably extracted from the lattice. This is easy to understand: in models with a smooth transition into the USR phase, the enhancement of scalar fluctuations primarily originates from a suppression of the inflaton velocity, as $\zeta \sim - H \delta\phi / \dot\phi$. The gradients of the inflaton field, which source GWs via $\partial_i\phi\,\partial_j\phi$, are not enhanced, and therefore the inflationary contribution to GWs remains negligible. 

\begin{figure*}
    \centering
    \includegraphics[width=\linewidth]{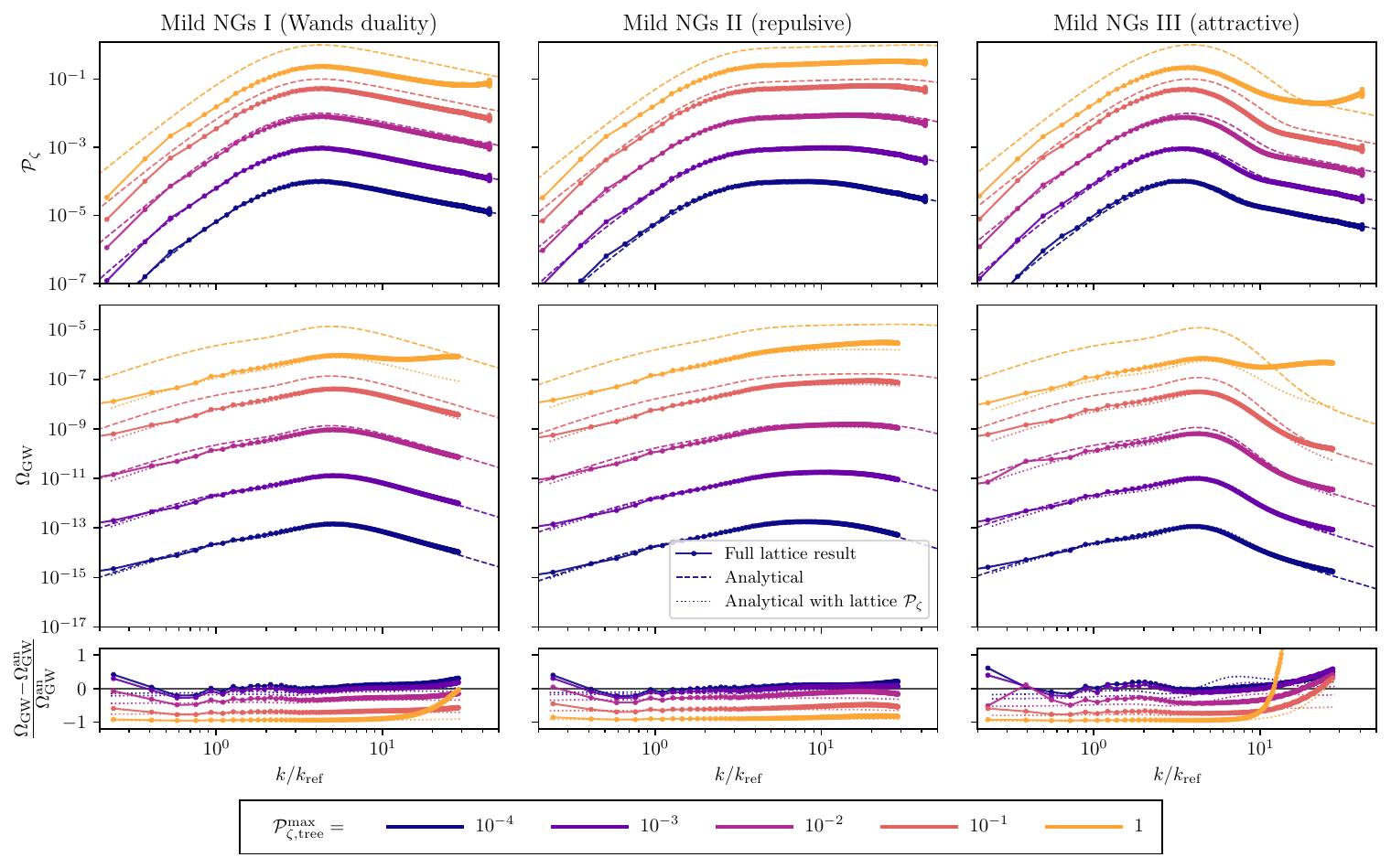}
    \caption{\textit{Top:} Lattice curvature power spectra from \cite{Caravano:2025diq} {(solid curve with circular marks)}, compared with linear perturbation theory {(dashed line)}. 
    \textit{Middle:} Scalar-induced GW spectra for the same models. 
    \textit{Bottom:} Fractional residuals with respect to the (semi-)analytical standard calculation. In all panels, different colours correspond to different values of the peak tree-level power spectrum, as indicated in the legend. 
    {In the middle and bottom panels, } solid curves with circular markers show the full lattice result, dashed curves show the standard semi-analytical prediction using the linear power spectrum, and dotted curves show the semi-analytical calculation using the lattice-derived curvature spectra from the top panel.}
    \label{fig:mild}
\end{figure*}

The post-reheating contribution, emitted at horizon re-entry, is of much greater observational relevance. We compute this component by evolving the post-reheating Newtonian potential as described in the previous section. The results for all 15 mildly non-Gaussian potentials are shown in Fig.~\ref{fig:mild}. The top panel shows the curvature power spectra computed in \cite{Caravano:2025diq}, the middle panel shows the resulting GW spectra, and the bottom panel shows the residuals relative to the standard semi-analytical calculation.

For small amplitudes of $\Omega_{\rm GW}$, the lattice results agree well with the standard semi-analytical prediction. At larger amplitudes, deviations appear and can reach $\mathcal{O}(1)$. To disentangle the origin of these differences, we also compute the semi-analytical prediction using the lattice power spectra (instead of the linear tree-level spectra) as input. These results correspond to the dotted curves in Fig.~\ref{fig:mild}. This allows us to separate the effect of non-Gaussianity from the effect of nonlinear inflationary dynamics and associated backreaction, which modify the curvature power spectrum itself. Differences between the dotted curve and the standard semi-analytic result are purely due to inflationary backreaction, while differences between the full lattice result and the dotted curve quantify the impact of non-Gaussianity.

From Fig.~\ref{fig:mild}, we learn that both effects, backreaction and non-Gaussianity, modify the standard prediction. The inflationary backreaction produces an overall shift in the GW spectrum. 
This shift originates from the correction to the inflaton velocity at the time 
the relevant modes exit the horizon during inflation. Inspecting the residuals 
more closely, we also see the impact of non-Gaussianity. As indicated by the 
difference between the full lattice result and the dotted curves, this effect 
is always present but predominantly affects the UV part of the spectrum. 
Moreover, except for the cases with $\mathcal{P}^{\rm max}_{\zeta} \sim 1$, 
the non-Gaussian contribution only induces $\mathcal{O}(1)$ corrections, 
in contrast with the inflationary backreaction, which can modify the GW signal 
by orders of magnitude in all these models. This reflects the fact that the 
non-Gaussianity generated in these potentials is mild.

We now comment on the differences between the individual cases. As discussed in \cite{Caravano:2024moy,Caravano:2025diq}, the three classes differ in their intrinsic non-Gaussianity. In case~I, $\delta\phi$ fluctuations remain almost perfectly Gaussian during inflation due to the Wands duality, and the resulting non-Gaussianity is entirely due to the nonlinear mapping between $\delta\phi$ and $\zeta$. In cases~II and~III, however, there is an additional source of non-Gaussianity: the intrinsic non-Gaussianity of $\delta\phi$ generated by a large $V^{(3)}(\phi)$ (and possibly higher derivatives) during the USR phase, which is absent in case~I. Inspecting the results, we see that cases~I and~II behave qualitatively similarly, with non-Gaussianity mostly affecting UV modes and the effect being slightly stronger in case~I, consistent with the repulsive non-Gaussianity reducing the effect of the nonlinear mapping, as we studied in \cite{Caravano:2025diq}. In contrast, case III---where a positive intrinsic non-Gaussianity is further amplified by the nonlinear mapping---exhibits non-Gaussian effects not only in the UV part of the GW spectrum but also in the IR. This is particularly evident in the residuals, where the dotted curve corresponding to the semi-analytical calculation using the lattice $\mathcal{P}_\zeta$ no longer matches the full lattice result, unlike in the other cases. In particular, we note that already for $\mathcal{P}^{\rm max}_{\zeta} \sim 10^{-4}$, a clear mismatch develops between the dotted curve and the full lattice GW spectrum, visible as a bump around $k \sim 10\,k_{\rm ref}$; this indicates that non-Gaussian effects already impact the spectrum.

These results demonstrate that intrinsic non-Gaussianity plays a significant role in shaping the final GW spectrum. As emphasized in \cite{Caravano:2024tlp, Caravano:2024moy}, intrinsic non-Gaussianity, particularly that generated prior to and around horizon crossing, can be difficult to capture in semi-analytical approaches based on the stochastic formalism, which rely on coarse-graining. Its inclusion is therefore a distinctive feature of nonlinear simulations. While more recent numerical implementations of stochastic inflation \cite{Mizuguchi:2024kbl,Launay:2024qsm,Launay:2025lnc} can in principle incorporate such effects, it would be interesting to compare how these contributions arise across the different frameworks. This becomes even more critical in the next section, where intrinsic nonlinearity and non-Gaussianity lead to substantially larger corrections to the observables.

Eventually, let us elaborate more on cases with large power spectra. As we discussed in Sec.~\ref{subsec:lattice-postinflation}, we do not expect substantial effects from nonlinear gravitational effects in the radiation era unless the gravitational potential becomes of order unity. This is particularly relevant in case~III, where, as discussed in \cite{Caravano:2025diq}, the pronounced UV growth of the curvature power
spectrum is associated with the trapping phenomenon:
regions that approach (but do not enter) the trapped regime can develop very
large curvature perturbations. However, these extreme regimes are not observationally relevant. Parameter regions in which trapping becomes strong
enough to substantially modify the GW spectrum are already excluded, as significant trapping leads to an excessive PBH abundance \cite{Caravano:2024tlp}. Thus, in observationally viable scenarios, trapping must remain a subdominant effect, and our linear evolution of the gravitational potential is reliable. These arguments are admittedly qualitative, and it would nevertheless be
valuable to investigate such effects in a quantitative manner using numerical-relativity and hydrodynamical simulations.

\subsection{Large non-Gaussianity}

\begin{figure}
    \centering
    \includegraphics[width=0.95\linewidth]{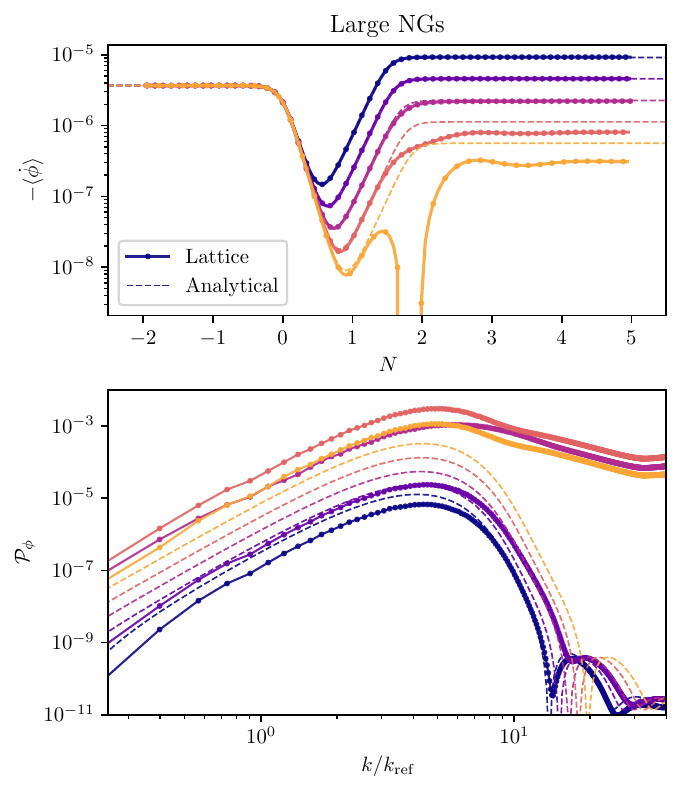}
    \caption{\textit{Top:} Evolution of the spatially averaged inflaton velocity in the large NG case, compared with the    prediction obtained from the background Klein--Gordon equation. 
    \textit{Bottom:} Inflaton power spectrum at the final simulation time compared with the linear prediction of the Mukhanov--Sasaki equation. 
    Colors correspond to the legend in Fig.~\ref{fig:mild}.}
    \label{fig:velocity_PS}
\end{figure}

\begin{figure}
    \centering
    \includegraphics[width=\linewidth]{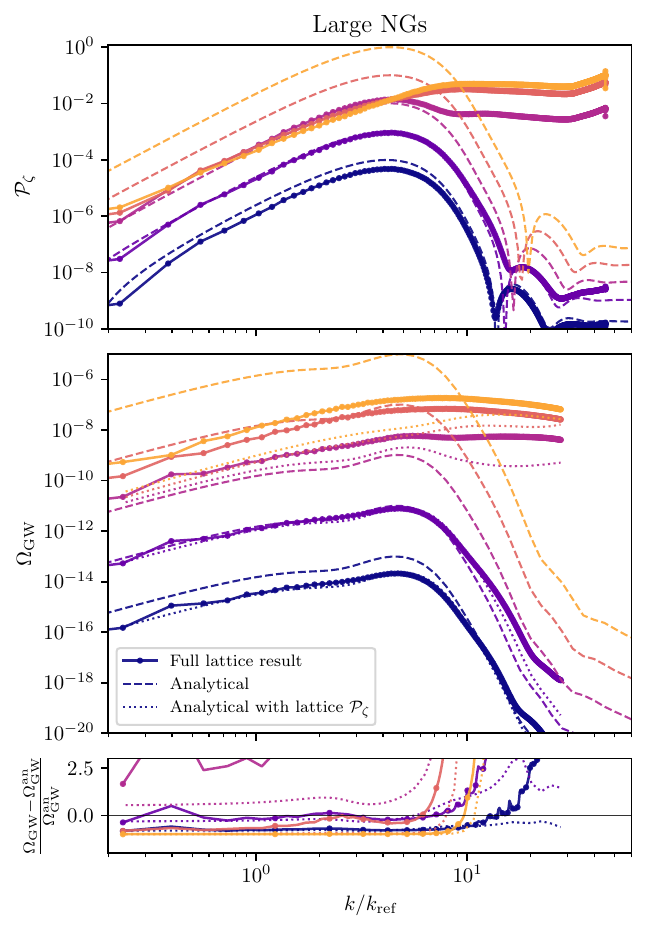}
    \caption{Same as Fig.~\ref{fig:mild}, but for the models with large non-Gaussianity presented in Sec.~\ref{large-NG}.}
    \label{fig:large_NG}
\end{figure}

We now turn to the cases characterised by large non-Gaussianity. As discussed in Sec.~\ref{sec:models}, these models feature both strong intrinsic non-Gaussianity, generated by the relatively sharp transition between the SR and USR phases, and a substantial nonlinear contribution from the mapping between $\delta\phi$ and $\zeta$, enhanced here by the large value of ${\eta_{H,{\rm III}}}$ during the second SR phase after USR.

\subsubsection{Inflationary dynamics and perturbations}
In contrast to the mild-NG cases presented previously, the inflationary dynamics of these large-NG models has not been studied prior to this work. We therefore begin by analysing the inflationary evolution, focusing on the ingredients needed to understand the subsequent behaviour of the GW signal. A more detailed characterisation of these models---including the structure of the non-Gaussianity and the trapping phenomenon---is deferred to Sec.~\ref{sec:trapping}.

The upper panel of Fig.~\ref{fig:velocity_PS} shows the evolution of the spatially averaged inflaton velocity across the lattice, compared with the standard perturbation theory background obtained from the Klein-Gordon equation. As expected, the velocity begins to deviate from the latter for large $\cal P_\zeta$, corresponding to lower values of the inflaton velocity in the USR phase. In the most extreme case, with $\mathcal{P}^{\rm max}_{\zeta,\rm tree}=1$, the averaged velocity even briefly flips sign. This behavior is a consequence of the trapping phenomenon taking place in many patches. Trapped regions of space roll back from the local maximum of the potential toward its minimum, instead of proceeding towards the second slow-roll attractor. This effect is visible in the evolution of the one-point probability density function (PDF) in Fig.~\ref{fig:PDF}, and even more so in the animation available at the bottom of the following \href{https://github.com/caravangelo/USR-on-the-lattice.git}{link}.

A qualitatively distinct and particularly striking feature of these large-NG models is the inflaton power spectrum at the final simulation time, shown in the lower panel of Fig.~\ref{fig:velocity_PS}. Here, the lattice results differ dramatically from the linear-theory prediction, not only in the most extreme case but across all cases, including those with relatively small fluctuations. More precisely, for $\mathcal{P}^{\rm max}_{\zeta,\rm tree} \leq 10^{-3}$, the IR part of the spectrum is significantly reduced by about one order of magnitude, while the UV part is almost unaffected. Additionally, we see the beginning of a plateau in the UV, which corresponds to modes crossing the Hubble sphere when the inflationary dynamics was back to SR (stage IV). Instead, for $\mathcal{P}^{\rm max}_{\zeta,\rm tree} \geq 10^{-2}$, the trapping phenomenon takes place and enhances the UV part of the spectrum by six orders of magnitude, giving rise to an almost plateau-type shape. The effect on the IR is also significant, with a substantial enhancement of the power spectrum, but whose size diminishes as the tree-level power spectrum increases.
These effects reflect the genuinely nonlinear dynamics of the inflaton: higher-order interactions feed back into the evolution of $\phi$, substantially modifying its power spectrum in a way that cannot be captured by linear perturbation theory. Since these nonlinear effects give rise to a rich and highly informative pattern of scalar non-Gaussianity, we provide a dedicated analysis in Sec.~\ref{sec:trapping}.

Turning now to the curvature perturbation, the upper panel of Fig.~\ref{fig:large_NG} shows the super-horizon $\mathcal{P}_{\zeta}$. Interestingly, for all models with $\mathcal{P}^{\rm max}_{\zeta,\rm tree}\geq 10^{-2}$, the IR part of the spectrum appears to approach a nearly universal shape. While we do not currently have a rigorous explanation for this behaviour, it may be intuitively related to the fact that, due to backreaction, these IR modes probe essentially the same region of the potential at similar times during inflation. As for the UV part of the spectrum, it inherits the huge enhancement and plateau-type structure of the inflaton spectrum for the three most extreme cases with trapping.

\subsubsection{Post-reheating GW spectrum}
In the large-NG cases, like in all other models considered in this work, the total GW signal is dominated by the post-reheating contribution generated at horizon re-entry. In this regime, however, and in contrast to the mild-NG cases, the component emitted during inflation is large enough to be resolved by our lattice simulation. The corresponding inflationary tensor spectrum is presented in Appendix~\ref{app:inflationary-gws}.

The middle and bottom panels of Fig.~\ref{fig:large_NG} show the post-reheating GW spectrum and the corresponding fractional residuals relative to the semi-analytical prediction. As before, we also show the semi-analytical result obtained by using the lattice-determined scalar power spectrum as input. Remarkably, in all large-NG cases---and regardless of the overall amplitude of the GW signal---the standard semi-analytical method fails to reproduce both the shape and the amplitude of the spectrum. For $\mathcal{P}^{\rm max}_{\zeta,\rm tree}=10^{-4}$, the discrepancy is entirely due to nonlinear corrections to the inflationary scalar dynamics, as demonstrated by the agreement between the lattice result and the semi-analytical computation that uses the lattice $\mathcal{P}_{\zeta}$. This also holds for $\mathcal{P}^{\rm max}_{\zeta,\rm tree}=10^{-3}$, but not in the UV part of the spectrum, where genuine non-Gaussian imprints on the SIGW become visible.
For larger amplitudes, primordial non-Gaussianity plays a dominant and qualitatively distinct role. In particular, we find that they lead to an enhancement of the GW spectrum by about one order of magnitude across all wavenumbers. However, these results should be interpreted with caution. As we discussed for the mild NG cases in Sec.~\ref{sec:results_mild}, the trapping phenomenon leads to localized regions around trapped patches where the gravitational potential exceeds unity, requiring more elaborate simulations to accurately compute the GW spectrum.

\section{Trapping and peak structure}
\label{sec:trapping}

\begin{figure*}
    \centering
    \includegraphics[width=\linewidth]{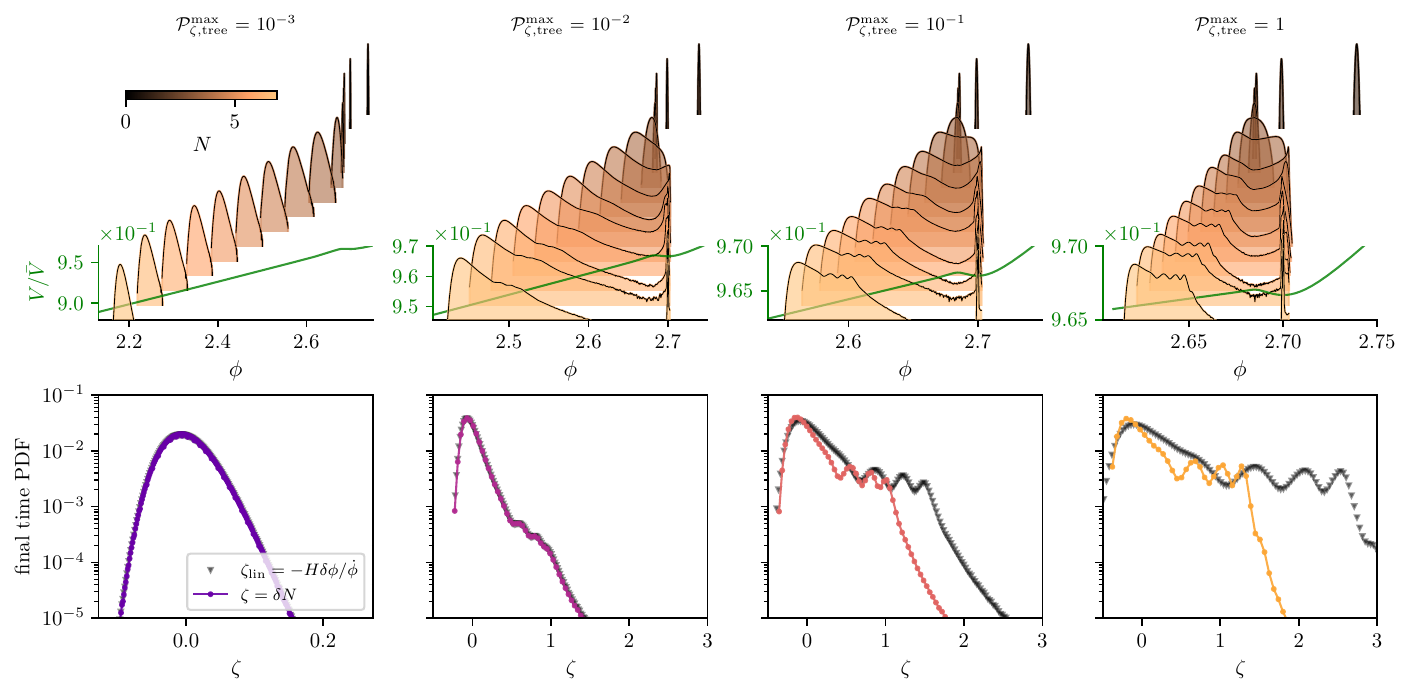}
    \caption{%
    \textit{Top:} One-point PDF of the inflaton field at different simulation times in the large-NG cases, as indicated by the colorbar in the top-left panel.  
    \textit{Bottom:} Final time PDF of $\zeta$, computed using both the linear relation between $\delta\phi$ and $\zeta$ and the {nonperturbative} $\delta N$ extraction (see~\cite{Caravano:2025diq,Caravano:2025klk} for details on the lattice procedure).  
    Different columns correspond to different values of the peak tree-level curvature power spectrum.}
    \label{fig:PDF}
\end{figure*}

Here we discuss additional aspects of the inflationary dynamics in the large-NG case, focusing on the trapping phenomenon and related non-Gaussian structures. In Fig.~\ref{fig:PDF}, we show the time evolution of the one-point PDF of the inflaton field during the simulation.

From these plots we observe that the final-time non-Gaussianity is significant in all cases. 
For the model with $\mathcal{P}^{\rm max}_{\zeta,\rm tree}=10^{-3}$, non-Gaussianity appears as a pronounced tail in the PDF of the inflaton field. In this case the dominant contribution arises from the intrinsic non-Gaussianity generated by the inflaton dynamics around horizon crossing, as indicated by the fact that applying the nonlinear $\delta N$ mapping does not significantly modify the PDF shape. A similar behavior persists for the case $\mathcal{P}^{\rm max}_{\zeta,\rm tree}=10^{-2}$ (second column), although oscillation-like features start to appear in the PDF of the inflaton. These oscillatory structures become more pronounced for the cases with $\mathcal{P}^{\rm max}_{\zeta,\rm tree}=10^{-1}$ and $1$. In these latter cases the nonlinear mapping between $\delta\phi$ and $\zeta$ significantly modifies the position and shape of the peaks in the PDF.

This oscillatory behavior is very similar to the one observed in~\cite{Caravano:2024tlp}, indicating that it is largely independent of the detailed shape of the inflationary potential, and is instead closely connected to the trapping phenomenon. Indeed, all cases with $\mathcal{P}^{\rm max}_{\zeta,\rm tree}\geq 10^{-2}$ exhibit trapping.

\begin{figure}
    \centering
    \includegraphics[width=0.9\linewidth]{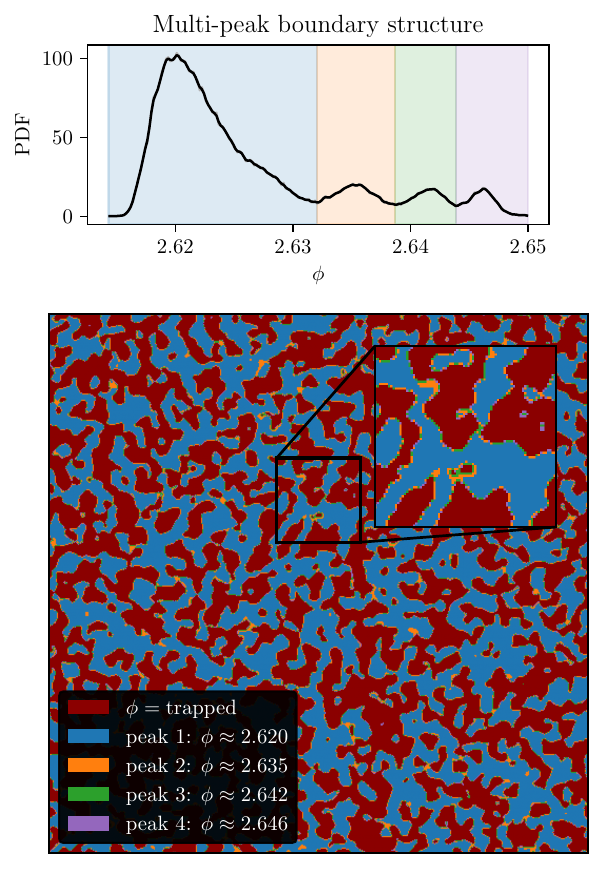}
    \caption{%
    Final-time simulation PDF (upper panel) and two-dimensional field snapshot (lower panel) for the large-NG case with $\mathcal{P}^{\rm max}_{\zeta,\rm tree}=1$.  
    Different colors indicate different regions of the PDF: red denotes trapped regions located in the local minimum of the potential, while other colors correspond to field values associated with different peaks in the PDF, as indicated in the legend.}
    \label{fig:PDF_snap}
\end{figure}

Given the universal emergence of these oscillatory features when the inflationary potential contains a local minimum, we now examine in more detail the field configuration associated to them. In Fig.~\ref{fig:PDF_snap}, we show a final-time snapshot of the simulation for the case $\mathcal{P}^{\rm max}_{\zeta,\rm tree}=1$. In this snapshot, different colors identify regions of the Universe with different dynamical behaviors. The red regions correspond to patches that become trapped in the local minimum of the potential. The blue regions represent the bulk of the inflationary fluctuations, far from the PDF tails. Other colors identify regions whose field values correspond to different peaks in the PDF tail.

This visualization makes it clear that the oscillatory features in the PDF arise because trapped regions are surrounded by non-trapped shells that continue slow-roll inflation, but with a delay relative to the bulk of the inflating Universe. By the end of the simulation (the time shown in the snapshot), these regions have effectively become causally disconnected from the bulk.

A deeper understanding of the physical origin of this multi-peak boundary structure around trapped surfaces requires further investigation. In this context, it would be valuable to compare our simulation results with previous studies of the trapping phenomenon~\cite{Garriga:2015fdk,Deng:2017uwc,Atal:2019cdz,Atal:2019erb,Escriva:2023uko}. However, it is interesting to speculate on the relation between this phenomenon and extreme value statistics. Indeed, one can interpret the phenomenon of overcoming the local maximum of the potential, and thus avoiding trapping, as a threshold effect: field trajectories that are sufficiently slowed down remain trapped, while those that retain enough velocity escape. Considering the distribution of maxima across a set of $N$ independent patches, one can describe it using extreme value statistics. For simplicity, assuming a Gaussian distribution of field values (which is certainly an approximation), the expectation values of the first-, second-, and higher-ranked peaks can be estimated. Following the interpretation of Ref.~\cite{MoradinezhadDizgah:2019wjf,Franciolini:2026jat}, these differently ranked peaks are on average separated by a distance proportional to the field variance divided by $\log(N)$ \cite{david2004order}. Estimating the variance from the width of the dominant peak in the $\phi$ distribution and taking $N$ as the number of independent patches, one obtains a separation scale roughly consistent with that observed in Fig.~\ref{fig:PDF_snap}.~\footnote{This multi-peak structure is not apparent in the low-$\mathcal{P}_\zeta$ cases, as trapping is more rare, so the corresponding peaks lie in the far tail of the distribution.}

Importantly, the quantity $N$ should not be confused with the number of lattice points used in the simulation, as increasing the resolution $N_{\rm pts}$ does not affect the location of the peaks. Rather, $N$ is the number of approximately independent regions in the simulation volume, i.e.~the number of patches of size comparable to the correlation length of the field fluctuations (set by the reference scale $k_{\rm ref}$).

\section{Conclusions}\label{sec:conclusions}

Gravitational waves have emerged as one of the most promising probes of early-Universe physics on scales far smaller than those accessible to traditional cosmological observations. Among the most compelling targets are stochastic backgrounds of scalar-induced gravitational waves, generated beyond linear perturbation theory by enhanced scalar perturbations, which can lie in the sensitivity bands of pulsar-timing arrays and future space- and ground-based interferometers. In many well-motivated scenarios, including models featuring a transient ultra-slow-roll (USR) phase, the scalar sector can become strongly amplified and highly non-Gaussian. In such regimes, it is important to assess the validity of the standard semi-analytical SIGW calculation, which relies on Gaussian statistics and linear perturbation theory.

In this work we addressed this question using lattice simulations. We developed a numerical pipeline that allows us to compute the SIGW background from first principles. In the first stage, we evolve the inflaton field nonlinearly through the SR--USR--SR transition and extract the super-horizon curvature perturbation $\zeta(\bm{x})$ nonperturbatively using the lattice $\delta N$ procedure developed in \cite{Caravano:2025diq}. In the second stage, we propagate these primordial perturbations into the post-reheating radiation-dominated era, evolving the Newtonian potential while retaining the full non-Gaussian structure of the primordial fluctuations in the tensor source. This approach allows us to compute the induced GW signal without assuming Gaussian statistics for the scalar perturbations, and provides a direct benchmark for the standard semi-analytical prediction.

Our results show that deviations from the standard semi-analytical SIGW calculation arise from two distinct effects: nonlinear corrections to the scalar dynamics during inflation, which modify the curvature power spectrum relative to linear perturbation theory, and genuine non-Gaussian primordial statistics, which affect the gravitational-wave signal beyond what is captured by the scalar power spectrum alone.
In the mildly non-Gaussian models considered here, the semi-analytical approach typically reproduces the correct order of magnitude of the GW signal, although it can receive sizable corrections, particularly in the UV part of the spectrum and for the largest amplitudes of $\mathcal{P}_\zeta$. For models with large non-Gaussianity, instead, the standard prediction can fail dramatically in both amplitude and spectral shape, independently of the overall magnitude of the GW signal, and in that case, the effect is largely driven by the nonlinear phenomena \textit{during} inflation.

These results highlight that reliable SIGW predictions require nonperturbative control of the scalar inflationary dynamics. In particular, our results show that interpreting a possible detection in terms of inflation requires theoretical predictions that properly account for nonlinear effects and primordial non-Gaussianities. For example, our findings indicate that it might be  more difficult to interpret the signal seen by pulsar timing arrays as originating from a SIGW background and  accompanied by a sizable PBH abundance \cite{Franciolini:2023pbf}.
The numerical pipeline developed in this work provides a controlled framework for performing such calculations. Since the code used in this study is \href{https://github.com/caravangelo/inflation-easy.git}{publicly available}, we hope that it will serve as a useful tool for the community to produce robust predictions for SIGWs and to assess theoretical uncertainties in different inflationary scenarios.

Our results provide a significant improvement in the way the SIGW background is calculated compared to the standard second-order computation, as it includes the primordial non-Gaussianities imprinted in the initial conditions in a fully nonperturbative way. At the same time, there remain several directions in which the framework can be further extended. First, although our results for the curvature power spectrum are nonlinear, the relation between the curvature perturbation and the gravitational potential (\ref{pert}) is treated at linear order; extending this relation beyond linear perturbation theory, along the lines of what is done on super-horizon scales in the matter-dominated era \cite{Bartolo:2005fp}, would be an important step. In addition, we have adopted a second-order source for the GWs. Progress in this direction has already been made in Ref.~\cite{Iovino:2024sgs}, and developing a fully consistent higher-order treatment would be a natural continuation of this work. Finally, as emphasized above, the gravitational potential is evolved linearly in the radiation era, an approximation that may require refinement when the gravitational potential itself becomes large.

Looking ahead, several extensions of the framework developed here would be worth pursuing. On the inflationary side, it would be interesting to explore a broader class of inflationary models and to further characterize how different sources of primordial non-Gaussianity propagate into the SIGW signal. On the post-reheating side, improving the treatment of the relation between $\zeta$ and $\Phi$, as well as incorporating higher-order contributions to the tensor source and to the gravitational evolution, would help further strengthen the connection between inflationary dynamics and observable gravitational-wave signals. Eventually, it would also be important to include a fully hydrodynamical description of the density perturbations, to go beyond the assumed linear order evolution within the sub-Hubble regime. These developments will become increasingly relevant as gravitational-wave observations continue to improve, making it essential to control theoretical uncertainties in SIGW predictions with a comparable level of precision.

\let\oldaddcontentsline\addcontentsline
\renewcommand{\addcontentsline}[3]{}

\section*{Data Availability}

The data underlying the results presented in this work are publicly available on Zenodo~\cite{Caravano:2026data}.
The repository includes the raw simulation outputs, the analytical benchmark files used in the comparisons, and a minimal Jupyter notebook to reproduce the main figures of the paper.  
The code used to generate these data is publicly available in the \texttt{InflationEasy} repository at \href{https://github.com/caravangelo/inflation-easy}{https://github.com/caravangelo/inflation-easy}.

\section*{Generative AI Statement}

The authors used generative AI tools to support proofreading of the manuscript and to help organize the publicly released data package described in the Data Availability section.  
All scientific analyses, methodological choices, results, and final text were reviewed and validated by the authors, who take full responsibility for the content of this work.

\begin{acknowledgments}
We thank Antonio Riotto for insightful discussions throughout the completion of this work, and insightful inputs on possible extensions of the techniques developed in this work. We are also grateful to Jaume Garriga for particularly illuminating discussions. We also thank Guillermo Ballesteros and Matteo Braglia for discussions on the inflationary SIGW emission, and Yoann Launay for comments on the draft.
AC acknowledges funding from the European Union's Horizon Europe research and innovation programme under the Marie Sk{\l}odowska-Curie grant agreement No. 101202657.
G.F.~acknowledges support by the
Italian MUR Departments of Excellence grant 2023--2027
``Quantum Frontier'' and from Istituto Nazionale di Fisica Nucleare (INFN) through the Theoretical Astroparticle Physics (TAsP) project.
\end{acknowledgments}

\appendix

\section{Inflationary gravitational waves}
\label{app:inflationary-gws}

In this Appendix, we discuss the contribution to the SIGW background arising from the emission of tensor modes around the epoch of Hubble crossing during inflation. Here, we limit ourselves to outlining the main ingredients of the computation which is then implemented in the lattice code, while leaving a systematic treatment to future work.

The source term for tensor modes is suppressed on super-Hubble scales, which implies that the dominant contribution during inflation is expected to originate from the near-horizon regime. The analysis, however, is more subtle than in the standard post-inflationary case, as it requires a careful treatment of the scalar perturbations while they are still in their quantum regime deep inside the Hubble radius. In particular, the naive application of classical perturbation theory is not sufficient, and the computation must be properly renormalized.
In what follows, we sketch the structure of the calculation and set an arbitrary cut-off, while leaving a suitable renormalization procedure, which removes spurious divergences and isolates the physical part of the signal, for future work. 

\subsubsection{Perturbation theory prediction}

In our setting, as done in the lattice code, we evaluate the source of tensor modes at second order during inflation in the flat gauge.  
In this gauge, the source takes the form
\begin{equation}
{\cal S}_{ij}^{\rm (infl)} 
= \partial_i\phi\,\partial_j\phi,
\label{eq:source_inflation}
\end{equation}
which captures the leading contribution to the anisotropic stress sourcing TT modes at second order. Additional terms arising from metric perturbations 
are suppressed by $\epsilon$ and are neglected here, consistent with the decoupling-limit treatment adopted in the rest of this work. For a general relativistic treatment, see \cite{Florio:2024pgm,Launay:2024qsm,Launay:2025kef,Launay:2025lnc}.

The power spectrum of curvature perturbations is obtained following the same procedure performed in Sec.~\ref{semi-analytical-GW-spectrum}, but adopting different kernel and transfer functions. Indeed, ${\cal P}_h$ takes the form \eqref{eq:P_h_ts}, with the kernel function being \cite{Fumagalli:2021mpc,Ballesteros:2022hjk} 
\begin{align}
\overline {I^2(t,s,k,\eta)} = &
16 
\bigg|
\frac{\sin(k\eta)}{k\eta}
\int_{\eta_h}^0 
k \d\eta'
\left( 1 - \frac{\mathcal{H}'(\eta ')}{\mathcal{H}^2(\eta ')} \right)
\nonumber 
\\
&\times
k F_{\rm inf}(0,\eta')
\frac{\zeta(ku ,\eta)}{\zeta(ku ,0)}\frac{\zeta(kv,\eta)}{\zeta(kv,0)}
\bigg|^2. 
\label{eq:kernel-inflationary-contribution}
\end{align}
We evaluate the power spectrum of inflationary tensor modes at the end of inflation, and thus the GW transfer function appearing as a prefactor in \eqref{eq:P_h_ts} becomes $T^h_k (\eta \to 0) =  \sin(k \eta)/(k \eta) =1$.
Finally, we introduced \cite{Ballesteros:2022hjk}
\begin{align}
k F_{\rm inf}(\eta,\eta')
&=-\frac{1}{(k\eta')^2}\Big[k(\eta-\eta')\cos\Big(k(\eta-\eta')\Big)
\nonumber \\
&-(1+k^2\eta\eta')\sin\Big(k(\eta-\eta')\Big)\Big].
\end{align}
It is simple to show that the integral is divergent when the initial time of emission $\eta_h \to - \infty$. 

In what follows, we assume $\eta_h$ to be the time at which all relevant $k$ modes are well inside the horizon, by a factor of $\mathcal{O}(\text{few})$. This choice arbitrarily makes the result convergent but is expected to provide a reliable order-of-magnitude estimate of the renormalized inflationary contribution. In the next section, where we present the lattice results, we focus exclusively on the SIGW spectrum in the large NG regime. In this case, the contribution from excited states---enhanced by the sharp transitions around the USR phase---is expected to dominate over the adiabatic component, see \cite{Fumagalli:2021mpc}. A careful derivation of the renormalized spectrum, as well as a consistent procedure to properly renormalize the (classical) lattice result, are left for future work.



\subsubsection{Numerical implementation}

During inflation we evolve a single canonical scalar field $\phi$ on a three-dimensional lattice using the fully nonlinear Klein--Gordon {equation coupled to the homogeneous Friedmann equations in conformal time}. The second-order tensor source appearing in Eq.~\eqref{eq:lattice-eom-hij} is the flat-gauge quadratic stress given in Eq.~\eqref{eq:source_inflation}, which captures the leading contribution to the anisotropic stress sourcing TT modes at second order. Additional terms arising from metric perturbations and constraint equations are suppressed by slow-roll parameters and are neglected here, consistent with the decoupling-limit treatment adopted in the rest of this work.

{Tensor modes are evolved without including their backreaction on the inflaton dynamics, while vector/shear modes are not included. These approximations are expected to remain valid as long as the energy stored in these modes is subdominant relative to the inflaton energy density. For tensor modes, this is verified a posteriori by the small inflationary GW contribution shown in Fig.~\ref{fig:large_NG_inflationary}, whereas shear backreaction may become relevant locally in the most extreme trapping regime considered in this work, where a fully relativistic treatment would be required.}

All fields are evolved with a staggered (leapfrog) integrator and periodic spatial boundary conditions. The output of this stage is a {nonperturbatively reconstructed} super-Hubble curvature perturbation $\zeta(\mathbf{x})$, obtained via the lattice-$\delta N$ procedure of Ref.~\cite{Caravano:2025diq}. This curvature field sets the initial condition for the post-reheating Newtonian potential.

\subsubsection{Simulation results}

\begin{figure}
    \centering
    \includegraphics[width=\linewidth]{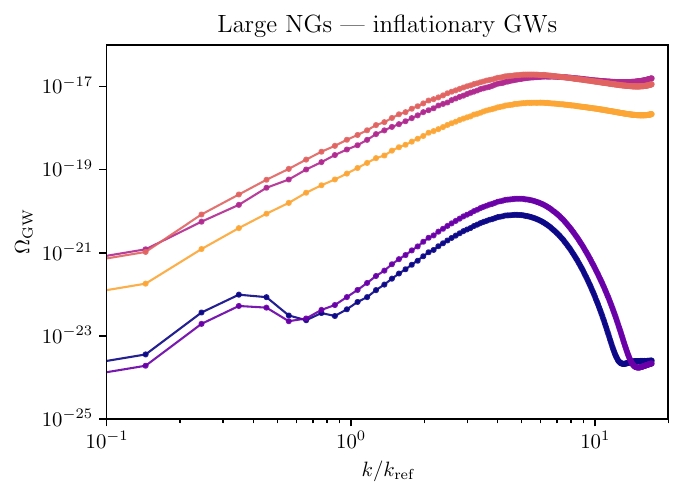}
    \caption{Inflationary GW power spectrum calculated from the lattice simulation in the Large NG setup.}
    \label{fig:large_NG_inflationary}
\end{figure}

As discussed earlier, the sharp SR--USR transition in these models also enhances scalar gradients during inflation. In contrast to the mild-NG cases, this makes the inflationary contribution to the GW signal non-negligible. Figure~\ref{fig:large_NG_inflationary} shows the inflationary GW power spectrum obtained directly from the lattice. Although still small and entirely subdominant compared to the post-reheating signal, its amplitude is now well above the numerical noise floor, allowing us to capture it.

Because this component remains very small in the present scenario, we do not pursue a deeper analysis here. In particular, we do not attempt to construct an analytical prediction for the inflationary GW spectrum, nor do we perform UV/IR convergence tests analogous to those carried out for the dominant post-reheating contribution. Both of these steps would require a careful treatment of the regularisation inherent to the inflationary GW calculation (especially on the analytical side) and are therefore beyond the scope of the present work.

A detailed study of the inflationary contribution is left for future work, where we will consider scenarios in which this component is expected to dominate the total GW signal, such as the models analyzed in~\cite{Caravano:2024tlp}.

\section{Convergence tests}
\begin{figure*}
    \centering
    \includegraphics[width=0.45\linewidth]{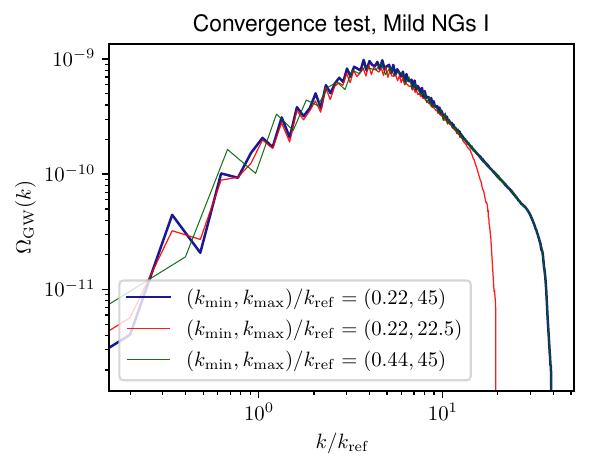}
    \includegraphics[width=0.45\linewidth]{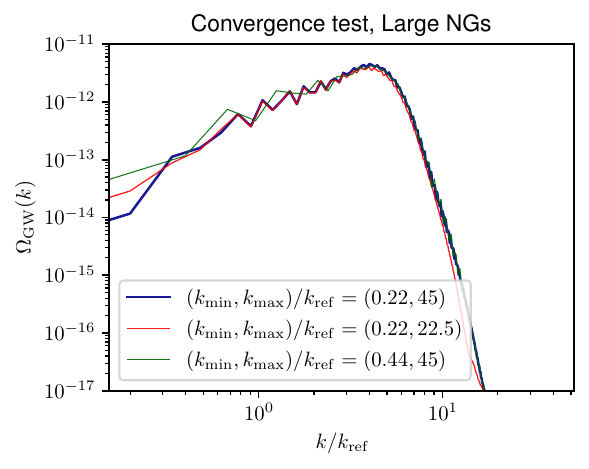}
   \caption{Convergence tests of the lattice simulation with respect to the spatial UV and IR resolutions. The left and right panels correspond to the mild and large NG cases, respectively. Here, the blue line represents the lattice setup used throughout the paper, while the other two lines are obtained by varying both the number of lattice points $N_{\rm pts}$ and the comoving box size $L$, thereby changing the UV and IR cutoffs.}
    \label{fig:convergence}
\end{figure*}

In this section, we present convergence tests of our lattice results. We consider both mild and large NG regimes. For each regime, we focus on the sub-case with the largest GW power spectrum that does not lead to dominant trapping. When trapping becomes a dominant phenomenon, (i) the model is not observationally relevant due to PBH overproduction, and (ii) the simulation does not yield reliable results for the sourced GW spectrum, as discussed at the end of Sec.~\ref{sec:results_mild}. As shown in Fig.~\ref{fig:convergence}, the results are highly stable: peak differences remain at the percent level even when both the IR and UV cutoffs are varied by factors of order two. This robustness ultimately reflects the convergence of both the inflationary and post-reheating stages of the simulation.

\begin{figure}
    \centering
\includegraphics[width=0.5\textwidth]{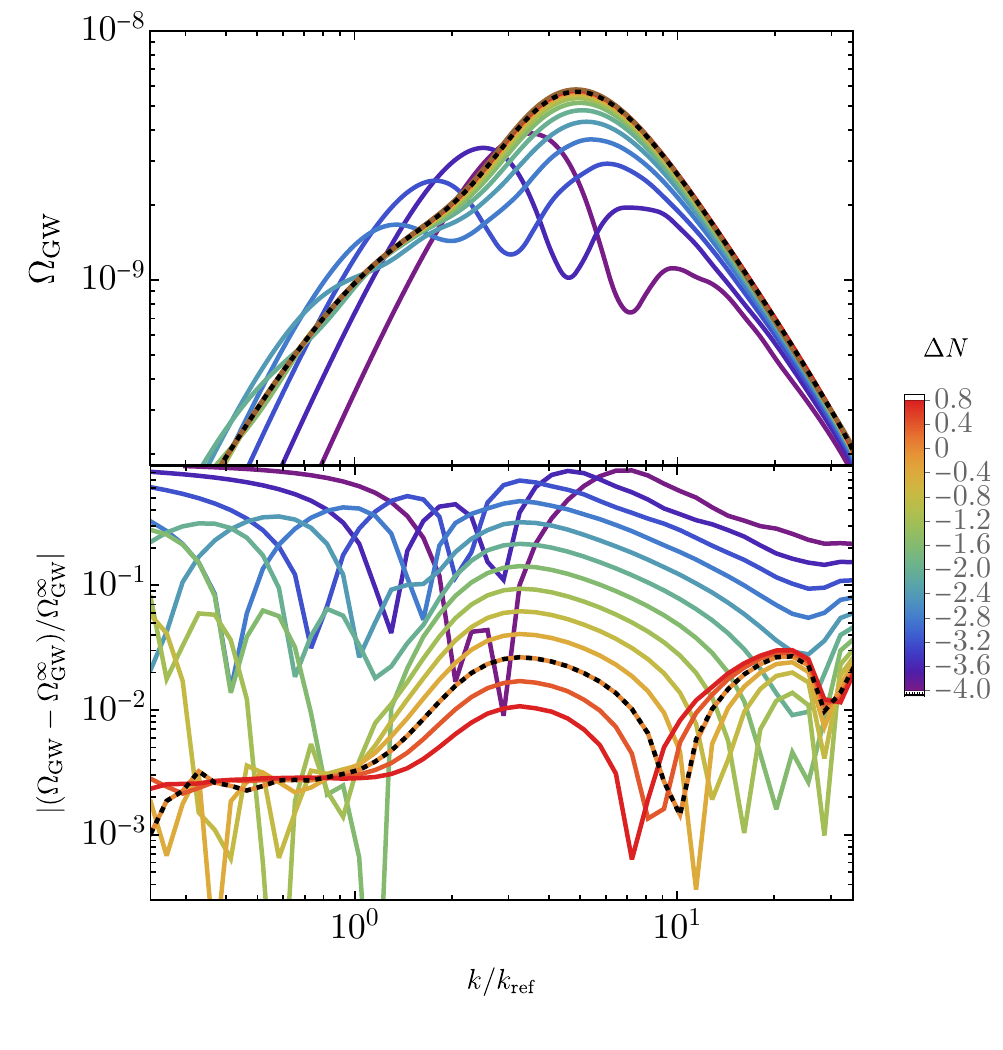}
    \caption{
Post-reheating GW spectrum as a function of $k$ for the case with ${\cal P}_\zeta^{\rm max, tree} = 10^{-4}$. Different colors indicate different evaluation times, tracked as the number of e-folds from the time where the lattice simulation stops, corresponding to $|k_{\rm max} \eta_{\rm end}^{\rm sim} |= 550$, following the semi-analytical approach based on the Gaussian approximation, Sec.~\ref{semi-analytical-GW-spectrum}.
Instead, the dashed line corresponds to the result obtained 
with the lattice simulation. In the bottom panel, we compare the spectra to its analytical asymptotic result in the limit $ k \eta \to \infty$. This shows that our choice of $\eta_{\rm end}^{\rm sim}$ captures most of the emission, with few percent error. 
Notice that $k_{\rm min} \eta_{\rm end}^{\rm sim} \simeq 2.8$.
This explains the large oscillations in the IR numerical spectra, which are not explained solely due to cosmic variance, as much larger than those observed in the curvature spectra. 
 }
    \label{fig:analytic_convergence}
\end{figure}

We also verified convergence with respect to both the initial and final simulation times. In practice, we checked that the GW spectra extracted from the simulation are effectively time-independent and do not depend on the chosen initial or final times. Convergence with respect to the final time can also be understood semi-analytically, as illustrated in Fig.~\ref{fig:analytic_convergence}, where we show the time convergence of the semi-analytical approach based on the Gaussian approximation. This analysis confirms that, by the end of the lattice simulations, the spectra have converged to their asymptotic values at the percent level.

\twocolumngrid
\bibliography{main}

@article{NANOGrav:2023gor,
  author =        {Agazie, Gabriella and others},
  journal =       {Astrophys. J. Lett.},
  number =        {1},
  pages =         {L8},
  title =         {{The NANOGrav 15 yr Data Set: Evidence for a
                   Gravitational-wave Background}},
  volume =        {951},
  year =          {2023},
  doi =           {10.3847/2041-8213/acdac6},
}

@article{EPTA:2023fyk,
  author =        {Antoniadis, J. and others},
  journal =       {Astron. Astrophys.},
  pages =         {A50},
  title =         {{The second data release from the European Pulsar
                   Timing Array - III. Search for gravitational wave
                   signals}},
  volume =        {678},
  year =          {2023},
  doi =           {10.1051/0004-6361/202346844},
}

@article{Sesana:2025udx,
  author =        {Sesana, Alberto and Figueroa, Daniel G.},
  journal =       {Ann. Rev. Astron. Astrophys.},
  pages =         {537--586},
  title =         {{Nanohertz Gravitational Waves}},
  volume =        {64},
  year =          {2026},
  doi =           {10.1146/annurev-astro-122325-025132},
}

@article{LISA:2024hlh,
  author =        {Colpi, Monica and others},
  archiveprefix = {arXiv},
  eprint =        {2402.07571},
  month =         {2},
  primaryclass =  {astro-ph.CO},
  reportnumber =  {ESA-SCI-DIR-RP-002},
  title =         {{LISA Definition Study Report}},
  year =          {2024},
}

@article{Kawamura:2020pcg,
  author =        {Kawamura, Seiji and others},
  journal =       {PTEP},
  number =        {5},
  pages =         {05A105},
  title =         {{Current status of space gravitational wave antenna
                   DECIGO and B-DECIGO}},
  volume =        {2021},
  year =          {2021},
  doi =           {10.1093/ptep/ptab019},
}

@article{Evans:2021gyd,
  author =        {Evans, Matthew and others},
  archiveprefix = {arXiv},
  eprint =        {2109.09882},
  month =         {9},
  primaryclass =  {astro-ph.IM},
  reportnumber =  {CE-P2100003-v7},
  title =         {{A Horizon Study for Cosmic Explorer: Science,
                   Observatories, and Community}},
  year =          {2021},
}

@article{ET:2019dnz,
  author =        {Maggiore, Michele and others},
  journal =       {JCAP},
  pages =         {050},
  title =         {{Science Case for the Einstein Telescope}},
  volume =        {03},
  year =          {2020},
  doi =           {10.1088/1475-7516/2020/03/050},
}

@article{ET:2025xjr,
  author =        {Abac, Adrian and others},
  journal =       {JCAP},
  pages =         {081},
  title =         {{The Science of the Einstein Telescope}},
  volume =        {03},
  year =          {2026},
  doi =           {10.1088/1475-7516/2026/03/081},
}

@article{Tomita:1975kj,
  author =        {Tomita, Kenji},
  journal =       {Prog. Theor. Phys.},
  pages =         {730},
  title =         {{Evolution of Irregularities in a Chaotic Early
                   Universe}},
  volume =        {54},
  year =          {1975},
  doi =           {10.1143/PTP.54.730},
}

@article{Matarrese:1992rp,
  author =        {Matarrese, Sabino and Pantano, Ornella and
                   Saez, Diego},
  journal =       {Phys. Rev. D},
  pages =         {1311--1323},
  title =         {{A General relativistic approach to the nonlinear
                   evolution of collisionless matter}},
  volume =        {47},
  year =          {1993},
  doi =           {10.1103/PhysRevD.47.1311},
}

@article{Matarrese:1993zf,
  author =        {Matarrese, Sabino and Pantano, Ornella and
                   Saez, Diego},
  journal =       {Phys. Rev. Lett.},
  pages =         {320--323},
  title =         {{General relativistic dynamics of irrotational dust:
                   Cosmological implications}},
  volume =        {72},
  year =          {1994},
  doi =           {10.1103/PhysRevLett.72.320},
}

@article{Matarrese:1997ay,
  author =        {Matarrese, Sabino and Mollerach, Silvia and
                   Bruni, Marco},
  journal =       {Phys. Rev. D},
  pages =         {043504},
  title =         {{Second order perturbations of the Einstein-de Sitter
                   universe}},
  volume =        {58},
  year =          {1998},
  doi =           {10.1103/PhysRevD.58.043504},
}

@article{Acquaviva:2002ud,
  author =        {Acquaviva, Viviana and Bartolo, Nicola and
                   Matarrese, Sabino and Riotto, Antonio},
  journal =       {Nucl. Phys. B},
  pages =         {119--148},
  title =         {{Second order cosmological perturbations from
                   inflation}},
  volume =        {667},
  year =          {2003},
  doi =           {10.1016/S0550-3213(03)00550-9},
}

@article{Mollerach:2003nq,
  author =        {Mollerach, Silvia and Harari, Diego and
                   Matarrese, Sabino},
  journal =       {Phys. Rev. D},
  pages =         {063002},
  title =         {{CMB polarization from secondary vector and tensor
                   modes}},
  volume =        {69},
  year =          {2004},
  doi =           {10.1103/PhysRevD.69.063002},
}

@article{Carbone:2004iv,
  author =        {Carbone, Carmelita and Matarrese, Sabino},
  journal =       {Phys. Rev. D},
  pages =         {043508},
  title =         {{A Unified treatment of cosmological perturbations
                   from super-horizon to small scales}},
  volume =        {71},
  year =          {2005},
  doi =           {10.1103/PhysRevD.71.043508},
}

@article{Ananda:2006af,
  author =        {Ananda, Kishore N. and Clarkson, Chris and
                   Wands, David},
  journal =       {Phys. Rev. D},
  pages =         {123518},
  title =         {{The Cosmological gravitational wave background from
                   primordial density perturbations}},
  volume =        {75},
  year =          {2007},
  doi =           {10.1103/PhysRevD.75.123518},
}

@article{Baumann:2007zm,
  author =        {Baumann, Daniel and Steinhardt, Paul J. and
                   Takahashi, Keitaro and Ichiki, Kiyotomo},
  journal =       {Phys. Rev. D},
  pages =         {084019},
  title =         {{Gravitational Wave Spectrum Induced by Primordial
                   Scalar Perturbations}},
  volume =        {76},
  year =          {2007},
  doi =           {10.1103/PhysRevD.76.084019},
}

@article{Domenech:2021ztg,
  author =        {Dom\`enech, Guillem},
  journal =       {Universe},
  number =        {11},
  pages =         {398},
  title =         {{Scalar Induced Gravitational Waves Review}},
  volume =        {7},
  year =          {2021},
  doi =           {10.3390/universe7110398},
}

@book{Byrnes:2025tji,
  editor =        {Byrnes, Christian and Franciolini, Gabriele and
                   Harada, Tomohiro and Pani, Paolo and Sasaki, Misao},
  publisher =     {Springer},
  series =        {Springer Series in Astrophysics and Cosmology},
  title =         {{Primordial Black Holes}},
  year =          {2025},
  doi =           {10.1007/978-981-97-8887-3},
  isbn =          {978-981--978886-6, 978-981--978889-7,
                   978-981--978887-3},
}

@article{Fumagalli:2021dtd,
  author =        {Fumagalli, Jacopo and Pieroni, Mauro and
                   Renaux-Petel, S{\'e}bastien and Witkowski, Lukas T.},
  journal =       {JCAP},
  number =        {07},
  pages =         {020},
  title =         {{Detecting primordial features with LISA}},
  volume =        {07},
  year =          {2022},
  doi =           {10.1088/1475-7516/2022/07/020},
}

@article{LISACosmologyWorkingGroup:2024hsc,
  author =        {Braglia, Matteo and others},
  journal =       {JCAP},
  pages =         {032},
  title =         {{Gravitational waves from inflation in LISA:
                   reconstruction pipeline and physics interpretation}},
  volume =        {11},
  year =          {2024},
  doi =           {10.1088/1475-7516/2024/11/032},
}

@article{LISACosmologyWorkingGroup:2025vdz,
  author =        {Gammal, Jonas El and others},
  journal =       {JCAP},
  pages =         {062},
  title =         {{Reconstructing primordial curvature perturbations
                   via scalar-induced gravitational waves with LISA}},
  volume =        {05},
  year =          {2025},
  doi =           {10.1088/1475-7516/2025/05/062},
}

@article{Espinosa:2018eve,
  author =        {Espinosa, Jos{\'e} Ram{\'o}n and Racco, Davide and
                   Riotto, Antonio},
  journal =       {JCAP},
  pages =         {012},
  title =         {{A Cosmological Signature of the SM Higgs
                   Instability: Gravitational Waves}},
  volume =        {09},
  year =          {2018},
  doi =           {10.1088/1475-7516/2018/09/012},
}

@article{Kohri:2018awv,
  author =        {Kohri, Kazunori and Terada, Takahiro},
  journal =       {Phys. Rev. D},
  number =        {12},
  pages =         {123532},
  title =         {{Semianalytic calculation of gravitational wave
                   spectrum nonlinearly induced from primordial
                   curvature perturbations}},
  volume =        {97},
  year =          {2018},
  doi =           {10.1103/PhysRevD.97.123532},
}

@article{Cai:2018dig,
  author =        {Cai, Rong-gen and Pi, Shi and Sasaki, Misao},
  journal =       {Phys. Rev. Lett.},
  number =        {20},
  pages =         {201101},
  title =         {{Gravitational Waves Induced by non-Gaussian Scalar
                   Perturbations}},
  volume =        {122},
  year =          {2019},
  doi =           {10.1103/PhysRevLett.122.201101},
}

@article{Unal:2018yaa,
  author =        {Unal, Caner},
  journal =       {Phys. Rev. D},
  number =        {4},
  pages =         {041301},
  title =         {{Imprints of Primordial Non-Gaussianity on
                   Gravitational Wave Spectrum}},
  volume =        {99},
  year =          {2019},
  doi =           {10.1103/PhysRevD.99.041301},
}

@article{Yuan:2020iwf,
  author =        {Yuan, Chen and Huang, Qing-Guo},
  journal =       {Phys. Lett. B},
  pages =         {136606},
  title =         {{Gravitational waves induced by the local-type
                   non-Gaussian curvature perturbations}},
  volume =        {821},
  year =          {2021},
  doi =           {10.1016/j.physletb.2021.136606},
}

@article{Atal:2021jyo,
  author =        {Atal, Vicente and Dom\`enech, Guillem},
  journal =       {JCAP},
  note =          {[Erratum: JCAP 10, E01 (2023)]},
  pages =         {001},
  title =         {{Probing non-Gaussianities with the high frequency
                   tail of induced gravitational waves}},
  volume =        {06},
  year =          {2021},
  doi =           {10.1088/1475-7516/2021/06/001},
}

@article{Adshead:2021hnm,
  author =        {Adshead, Peter and Lozanov, Kaloian D. and
                   Weiner, Zachary J.},
  journal =       {JCAP},
  pages =         {080},
  title =         {{Non-Gaussianity and the induced gravitational wave
                   background}},
  volume =        {10},
  year =          {2021},
  doi =           {10.1088/1475-7516/2021/10/080},
}

@article{Garcia-Saenz:2022tzu,
  author =        {Garcia-Saenz, Sebastian and Pinol, Lucas and
                   Renaux-Petel, S\'ebastien and Werth, Denis},
  journal =       {JCAP},
  pages =         {057},
  title =         {{No-go theorem for scalar-trispectrum-induced
                   gravitational waves}},
  volume =        {03},
  year =          {2023},
  doi =           {10.1088/1475-7516/2023/03/057},
}

@article{Yuan:2023ofl,
  author =        {Yuan, Chen and Meng, De-Shuang and Huang, Qing-Guo},
  journal =       {JCAP},
  pages =         {036},
  title =         {{Full analysis of the scalar-induced gravitational
                   waves for the curvature perturbation with local-type
                   non-Gaussianities}},
  volume =        {12},
  year =          {2023},
  doi =           {10.1088/1475-7516/2023/12/036},
}

@article{Li:2023qua,
  author =        {Li, Jun-Peng and Wang, Sai and Zhao, Zhi-Chao and
                   Kohri, Kazunori},
  journal =       {JCAP},
  pages =         {056},
  title =         {{Primordial non-Gaussianity f $_{NL}$ and
                   anisotropies in scalar-induced gravitational waves}},
  volume =        {10},
  year =          {2023},
  doi =           {10.1088/1475-7516/2023/10/056},
}

@article{Li:2023xtl,
  author =        {Li, Jun-Peng and Wang, Sai and Zhao, Zhi-Chao and
                   Kohri, Kazunori},
  journal =       {JCAP},
  pages =         {039},
  title =         {{Complete analysis of the background and anisotropies
                   of scalar-induced gravitational waves: primordial
                   non-Gaussianity f $_{NL}$ and g $_{NL}$ considered}},
  volume =        {06},
  year =          {2024},
  doi =           {10.1088/1475-7516/2024/06/039},
}

@article{Perna:2024ehx,
  author =        {Perna, Gabriele and Testini, Chiara and
                   Ricciardone, Angelo and Matarrese, Sabino},
  journal =       {JCAP},
  pages =         {086},
  title =         {{Fully non-Gaussian Scalar-Induced Gravitational
                   Waves}},
  volume =        {05},
  year =          {2024},
  doi =           {10.1088/1475-7516/2024/05/086},
}

@article{Ruiz:2024weh,
  author =        {Ruiz, Juan \'Alvarez and Rey, Juli\'an},
  journal =       {JCAP},
  pages =         {026},
  title =         {{Gravitational waves in ultra-slow-roll and their
                   anisotropy at two loops}},
  volume =        {04},
  year =          {2025},
  doi =           {10.1088/1475-7516/2025/04/026},
}

@article{Li:2025met,
  author =        {Li, Jun-Peng and Wang, Sai and Zhao, Zhi-Chao and
                   Kohri, Kazunori},
  journal =       {JCAP},
  pages =         {064},
  title =         {{Isotropy, anisotropies and non-Gaussianity in the
                   scalar-induced gravitational-wave background:
                   diagrammatic approach for primordial non-Gaussianity
                   up to arbitrary order}},
  volume =        {05},
  year =          {2026},
  doi =           {10.1088/1475-7516/2026/05/064},
}

@article{Ferrante:2022mui,
  author =        {Ferrante, Giacomo and Franciolini, Gabriele and
                   Iovino, Junior., Antonio and Urbano, Alfredo},
  journal =       {Phys. Rev. D},
  number =        {4},
  pages =         {043520},
  title =         {{Primordial non-Gaussianity up to all orders:
                   Theoretical aspects and implications for primordial
                   black hole models}},
  volume =        {107},
  year =          {2023},
  doi =           {10.1103/PhysRevD.107.043520},
}

@article{Caravano:2024moy,
  author =        {Caravano, Angelo and Franciolini, Gabriele and
                   Renaux-Petel, S{\'e}bastien},
  journal =       {Phys. Rev. D},
  number =        {6},
  pages =         {063518},
  title =         {{Ultraslow-roll inflation on the lattice:
                   Backreaction and nonlinear effects}},
  volume =        {111},
  year =          {2025},
  doi =           {10.1103/PhysRevD.111.063518},
}

@article{Caravano:2025diq,
  author =        {Caravano, Angelo and Franciolini, Gabriele and
                   Renaux-Petel, S{\'e}bastien},
  journal =       {Phys. Rev. D},
  number =        {8},
  pages =         {083508},
  title =         {{Ultraslow-roll inflation on the lattice. II.
                   Nonperturbative curvature perturbation}},
  volume =        {112},
  year =          {2025},
  doi =           {10.1103/39qd-gdfm},
}

@article{Caravano:2025klk,
  author =        {Caravano, Angelo},
  archiveprefix = {arXiv},
  eprint =        {2506.11797},
  month =         {6},
  primaryclass =  {astro-ph.CO},
  title =         {{InflationEasy: A C++ Lattice Code for Inflation}},
  year =          {2025},
}

@article{Maggiore:1999vm,
  author =        {Maggiore, Michele},
  journal =       {Phys. Rept.},
  pages =         {283--367},
  title =         {{Gravitational wave experiments and early universe
                   cosmology}},
  volume =        {331},
  year =          {2000},
  doi =           {10.1016/S0370-1573(99)00102-7},
}

@article{Malik:2008im,
  author =        {Malik, Karim A. and Wands, David},
  journal =       {Phys. Rept.},
  pages =         {1--51},
  title =         {{Cosmological perturbations}},
  volume =        {475},
  year =          {2009},
  doi =           {10.1016/j.physrep.2009.03.001},
}

@article{DeLuca:2019ufz,
  author =        {De Luca, V. and Franciolini, G. and Kehagias, A. and
                   Riotto, A.},
  journal =       {JCAP},
  pages =         {014},
  title =         {{On the Gauge Invariance of Cosmological
                   Gravitational Waves}},
  volume =        {03},
  year =          {2020},
  doi =           {10.1088/1475-7516/2020/03/014},
}

@article{Inomata:2019yww,
  author =        {Inomata, Keisuke and Terada, Takahiro},
  journal =       {Phys. Rev. D},
  number =        {2},
  pages =         {023523},
  title =         {{Gauge Independence of Induced Gravitational Waves}},
  volume =        {101},
  year =          {2020},
  doi =           {10.1103/PhysRevD.101.023523},
}

@article{Yuan:2019fwv,
  author =        {Yuan, Chen and Chen, Zu-Cheng and Huang, Qing-Guo},
  journal =       {Phys. Rev. D},
  number =        {6},
  pages =         {063018},
  title =         {{Scalar Induced Gravitational Waves in Different
                   Gauges}},
  volume =        {101},
  year =          {2020},
  doi =           {10.1103/PhysRevD.101.063018},
}

@article{Domenech:2020xin,
  author =        {Dom{\`e}nech, Guillem and Sasaki, Misao},
  journal =       {Phys. Rev. D},
  number =        {6},
  pages =         {063531},
  title =         {{Approximate gauge independence of the induced
                   gravitational wave spectrum}},
  volume =        {103},
  year =          {2021},
  doi =           {10.1103/PhysRevD.103.063531},
}

@article{Domenech:2025ccu,
  author =        {Dom{\`e}nech, Guillem and Pi, Shi and Wang, Ao},
  journal =       {Phys. Rev. Lett.},
  number =        {22},
  pages =         {221402},
  title =         {{Observable Gravitational Wave Strain at Second
                   Order}},
  volume =        {136},
  year =          {2026},
  doi =           {10.1103/pwbs-xwrh},
}

@article{Garcia-Bellido:2007fiu,
  author =        {Garcia-Bellido, Juan and Figueroa, Daniel G. and
                   Sastre, Alfonso},
  journal =       {Phys. Rev. D},
  pages =         {043517},
  title =         {{A Gravitational Wave Background from Reheating after
                   Hybrid Inflation}},
  volume =        {77},
  year =          {2008},
  doi =           {10.1103/PhysRevD.77.043517},
}

@article{Caravano:2021pgc,
  author =        {Caravano, Angelo and Komatsu, Eiichiro and
                   Lozanov, Kaloian D. and Weller, Jochen},
  journal =       {JCAP},
  number =        {12},
  pages =         {010},
  title =         {{Lattice simulations of inflation}},
  volume =        {12},
  year =          {2021},
  doi =           {10.1088/1475-7516/2021/12/010},
}

@phdthesis{Caravano:2022yyv,
  author =        {Caravano, Angelo},
  month =         {7},
  school =        {Munich U.},
  title =         {{Simulating the inflationary Universe: from
                   single-field to the axion-U(1) model}},
  year =          {2022},
  doi =           {10.5282/edoc.30905},
}

@article{Zeng:2025cer,
  author =        {Zeng, Xiang-Xi and Ning, Zhuan and Cai, Rong-Gen and
                   Wang, Shao-Jiang},
  archiveprefix = {arXiv},
  eprint =        {2508.10812},
  month =         {8},
  primaryclass =  {astro-ph.CO},
  title =         {{Scalar-induced gravitational waves with
                   non-Gaussianity up to all orders}},
  year =          {2025},
}

@article{DeLuca:2023tun,
  author =        {De Luca, Valerio and Kehagias, Alex and
                   Riotto, Antonio},
  journal =       {Phys. Rev. D},
  number =        {6},
  pages =         {063531},
  title =         {{How well do we know the primordial black hole
                   abundance: The crucial role of nonlinearities when
                   approaching the horizon}},
  volume =        {108},
  year =          {2023},
  doi =           {10.1103/PhysRevD.108.063531},
}

@article{Ning:2025yvj,
  author =        {Ning, Zhuan and Yuwen, Zi-Yan and Zeng, Xiang-Xi and
                   Cai, Rong-Gen and Wang, Shao-Jiang},
  archiveprefix = {arXiv},
  eprint =        {2512.21151},
  month =         {12},
  primaryclass =  {gr-qc},
  title =         {{Acoustic gravitational waves from primordial
                   curvature perturbations}},
  year =          {2025},
}

@article{Franciolini:2022pav,
  author =        {Franciolini, Gabriele and Urbano, Alfredo},
  journal =       {Phys. Rev. D},
  number =        {12},
  pages =         {123519},
  title =         {{Primordial black hole dark matter from inflation:
                   The reverse engineering approach}},
  volume =        {106},
  year =          {2022},
  doi =           {10.1103/PhysRevD.106.123519},
}

@article{Byrnes:2018txb,
  author =        {Byrnes, Christian T. and Cole, Philippa S. and
                   Patil, Subodh P.},
  journal =       {JCAP},
  pages =         {028},
  title =         {{Steepest growth of the power spectrum and primordial
                   black holes}},
  volume =        {06},
  year =          {2019},
  doi =           {10.1088/1475-7516/2019/06/028},
}

@article{Taoso:2021uvl,
  author =        {Taoso, Marco and Urbano, Alfredo},
  journal =       {JCAP},
  pages =         {016},
  title =         {{Non-gaussianities for primordial black hole
                   formation}},
  volume =        {08},
  year =          {2021},
  doi =           {10.1088/1475-7516/2021/08/016},
}

@article{Cai:2018dkf,
  author =        {Cai, Yi-Fu and Chen, Xingang and
                   Namjoo, Mohammad Hossein and Sasaki, Misao and
                   Wang, Dong-Gang and Wang, Ziwei},
  journal =       {JCAP},
  pages =         {012},
  title =         {{Revisiting non-Gaussianity from non-attractor
                   inflation models}},
  volume =        {05},
  year =          {2018},
  doi =           {10.1088/1475-7516/2018/05/012},
}

@article{Atal:2019cdz,
  author =        {Atal, Vicente and Garriga, Jaume and
                   Marcos-Caballero, Airam},
  journal =       {JCAP},
  pages =         {073},
  title =         {{Primordial black hole formation with non-Gaussian
                   curvature perturbations}},
  volume =        {09},
  year =          {2019},
  doi =           {10.1088/1475-7516/2019/09/073},
}

@article{Biagetti:2021eep,
  author =        {Biagetti, Matteo and De Luca, Valerio and
                   Franciolini, Gabriele and Kehagias, Alex and
                   Riotto, Antonio},
  journal =       {Phys. Lett. B},
  pages =         {136602},
  title =         {{The formation probability of primordial black
                   holes}},
  volume =        {820},
  year =          {2021},
  doi =           {10.1016/j.physletb.2021.136602},
}

@article{Pi:2022ysn,
  author =        {Pi, Shi and Sasaki, Misao},
  journal =       {Phys. Rev. Lett.},
  number =        {1},
  pages =         {011002},
  title =         {{Logarithmic Duality of the Curvature Perturbation}},
  volume =        {131},
  year =          {2023},
  doi =           {10.1103/PhysRevLett.131.011002},
}

@article{Ballesteros:2024pwn,
  author =        {Ballesteros, Guillermo and Konstandin, Thomas and
                   P{\'e}rez Rodr{\'\i}guez, Alejandro and
                   Pierre, Mathias and Rey, Juli{\'a}n},
  journal =       {JCAP},
  pages =         {013},
  title =         {{Non-Gaussian tails without stochastic inflation}},
  volume =        {11},
  year =          {2024},
  doi =           {10.1088/1475-7516/2024/11/013},
}

@article{Inui:2024fgk,
  author =        {Inui, Ryoto and Joana, Cristian and Motohashi, Hayato and
                   Pi, Shi and Tada, Yuichiro and Yokoyama, Shuichiro},
  journal =       {JCAP},
  pages =         {021},
  title =         {{Primordial black holes and induced gravitational
                   waves from logarithmic non-Gaussianity}},
  volume =        {03},
  year =          {2025},
  doi =           {10.1088/1475-7516/2025/03/021},
}

@article{Wands:1998yp,
  author =        {Wands, David},
  journal =       {Phys. Rev. D},
  pages =         {023507},
  title =         {{Duality invariance of cosmological perturbation
                   spectra}},
  volume =        {60},
  year =          {1999},
  doi =           {10.1103/PhysRevD.60.023507},
}

@article{Caravano:2024tlp,
  author =        {Caravano, Angelo and Inomata, Keisuke and
                   Renaux-Petel, S\'ebastien},
  journal =       {Phys. Rev. Lett.},
  number =        {15},
  pages =         {151001},
  title =         {{Inflationary Butterfly Effect: Nonperturbative
                   Dynamics from Small-Scale Features}},
  volume =        {133},
  year =          {2024},
  doi =           {10.1103/PhysRevLett.133.151001},
}

@article{Mizuguchi:2024kbl,
  author =        {Mizuguchi, Yurino and Murata, Tomoaki and
                   Tada, Yuichiro},
  journal =       {JCAP},
  pages =         {050},
  reportnumber =  {RUP-24-10},
  title =         {{STOLAS: STOchastic LAttice Simulation of cosmic
                   inflation}},
  volume =        {12},
  year =          {2024},
  doi =           {10.1088/1475-7516/2024/12/050},
}

@article{Launay:2024qsm,
  author =        {Launay, Yoann L. and Rigopoulos, Gerasimos I. and
                   Shellard, E. Paul S.},
  journal =       {Phys. Rev. D},
  number =        {12},
  pages =         {123523},
  title =         {{Stochastic inflation in general relativity}},
  volume =        {109},
  year =          {2024},
  doi =           {10.1103/PhysRevD.109.123523},
}

@article{Launay:2025lnc,
  author =        {Launay, Yoann L. and Rigopoulos, Gerasimos I. and
                   Shellard, E. Paul S.},
  journal =       {Phys. Rev. D},
  number =        {12},
  pages =         {123515},
  title =         {{Stochastic Inflation in Numerical Relativity}},
  volume =        {113},
  year =          {2026},
  doi =           {10.1103/h2cb-q1mt},
}

@article{Garriga:2015fdk,
  author =        {Garriga, Jaume and Vilenkin, Alexander and
                   Zhang, Jun},
  journal =       {JCAP},
  pages =         {064},
  title =         {{Black holes and the multiverse}},
  volume =        {02},
  year =          {2016},
  doi =           {10.1088/1475-7516/2016/02/064},
}

@article{Deng:2017uwc,
  author =        {Deng, Heling and Vilenkin, Alexander},
  journal =       {JCAP},
  pages =         {044},
  title =         {{Primordial black hole formation by vacuum bubbles}},
  volume =        {12},
  year =          {2017},
  doi =           {10.1088/1475-7516/2017/12/044},
}

@article{Atal:2019erb,
  author =        {Atal, Vicente and Cid, Judith and Escriv\`a, Albert and
                   Garriga, Jaume},
  journal =       {JCAP},
  pages =         {022},
  title =         {{PBH in single field inflation: the effect of shape
                   dispersion and non-Gaussianities}},
  volume =        {05},
  year =          {2020},
  doi =           {10.1088/1475-7516/2020/05/022},
}

@article{Escriva:2023uko,
  author =        {Escriv\`a, Albert and Atal, Vicente and
                   Garriga, Jaume},
  journal =       {JCAP},
  pages =         {035},
  title =         {{Formation of trapped vacuum bubbles during
                   inflation, and consequences for PBH scenarios}},
  volume =        {10},
  year =          {2023},
  doi =           {10.1088/1475-7516/2023/10/035},
}

@article{MoradinezhadDizgah:2019wjf,
  author =        {Moradinezhad Dizgah, Azadeh and Franciolini, Gabriele and
                   Riotto, Antonio},
  journal =       {JCAP},
  pages =         {001},
  title =         {{Primordial Black Holes from Broad Spectra: Abundance
                   and Clustering}},
  volume =        {11},
  year =          {2019},
  doi =           {10.1088/1475-7516/2019/11/001},
}

@article{Franciolini:2026jat,
  author =        {Franciolini, G. and Kehagias, A. and Riotto, A.},
  journal =       {JCAP},
  pages =         {014},
  title =         {{Standard Model Higgs Peaks: a Note on the Vacuum
                   Instability during Inflation}},
  volume =        {07},
  year =          {2026},
  doi =           {10.1088/1475-7516/2026/07/014},
}

@book{david2004order,
  author =        {David, H.A. and Nagaraja, H.N.},
  publisher =     {Wiley},
  series =        {Wiley Series in Probability and Statistics},
  title =         {Order Statistics},
  year =          {2004},
  isbn =          {9780471654018},
  url =           {https://books.google.it/books?id=bdhzFXg6xFkC},
}

@article{Franciolini:2023pbf,
  author =        {Franciolini, Gabriele and Iovino, Junior., Antonio and
                   Vaskonen, Ville and Veermae, Hardi},
  journal =       {Phys. Rev. Lett.},
  number =        {20},
  pages =         {201401},
  title =         {{Recent Gravitational Wave Observation by Pulsar
                   Timing Arrays and Primordial Black Holes: The
                   Importance of Non-Gaussianities}},
  volume =        {131},
  year =          {2023},
  doi =           {10.1103/PhysRevLett.131.201401},
}

@article{Bartolo:2005fp,
  author =        {Bartolo, Nicola and Matarrese, Sabino and
                   Riotto, Antonio},
  journal =       {JCAP},
  pages =         {010},
  title =         {{Non-Gaussianity of Large-Scale Cosmic Microwave
                   Background Anisotropies beyond Perturbation Theory}},
  volume =        {08},
  year =          {2005},
  doi =           {10.1088/1475-7516/2005/08/010},
}

@article{Iovino:2024sgs,
  author =        {Iovino, A. J. and Matarrese, S. and Perna, G. and
                   Ricciardone, A. and Riotto, A.},
  journal =       {Phys. Lett. B},
  pages =         {140039},
  title =         {{How well do we know the scalar-induced gravitational
                   waves?}},
  volume =        {872},
  year =          {2026},
  doi =           {10.1016/j.physletb.2025.140039},
}

@misc{Caravano:2026data,
  author =        {Caravano, Angelo and Franciolini, Gabriele and
                   Renaux-Petel, S{\'e}bastien},
  howpublished =  {Zenodo},
  title =         {{Data and plotting notebook for ``Lattice simulations
                   of scalar-induced gravitational waves from
                   inflation''}},
  year =          {2026},
  doi =           {10.5281/zenodo.19387301},
}

@article{Florio:2024pgm,
  author =        {Florio, Ericka and Shellard, E. Paul S.},
  journal =       {Phys. Rev. D},
  number =        {2},
  pages =         {023534},
  title =         {{Fully relativistic evolution of vacuum tensor
                   inhomogeneities during inflation}},
  volume =        {113},
  year =          {2026},
  doi =           {10.1103/l48w-z27g},
}

@article{Launay:2025kef,
  author =        {Launay, Yoann L. and Rigopoulos, Gerasimos I. and
                   Shellard, E. Paul S.},
  journal =       {Phys. Rev. D},
  number =        {4},
  pages =         {043518},
  title =         {{Bunch-Davies initial conditions and nonperturbative
                   inflationary dynamics in numerical relativity}},
  volume =        {112},
  year =          {2025},
  doi =           {10.1103/qg66-gxh2},
}

@article{Fumagalli:2021mpc,
  author =        {Fumagalli, Jacopo and Palma, Gonzalo A. and
                   Renaux-Petel, S{\'e}bastien and Sypsas, Spyros and
                   Witkowski, Lukas T. and Zenteno, Cristobal},
  journal =       {JHEP},
  pages =         {196},
  title =         {{Primordial gravitational waves from excited states}},
  volume =        {03},
  year =          {2022},
  doi =           {10.1007/JHEP03(2022)196},
}

@article{Ballesteros:2022hjk,
  author =        {Ballesteros, Guillermo and Garc{\'\i}a, Marcos A. G. and
                   Rodr{\'\i}guez, Alejandro P{\'e}rez and
                   Pierre, Mathias and Rey, Juli{\'a}n},
  journal =       {JCAP},
  pages =         {006},
  title =         {{Primordial black holes and gravitational waves from
                   dissipation during inflation}},
  volume =        {12},
  year =          {2022},
  doi =           {10.1088/1475-7516/2022/12/006},
}
\let\addcontentsline\oldaddcontentsline

\end{document}